\documentclass{article}%
\usepackage{graphicx}
\usepackage{amsmath}
\usepackage{amsfonts}
\usepackage{amssymb}%
\providecommand{\U}[1]{\protect\rule{.1in}{.1in}}
\begin{document}

\author{Antony Valentini\\Augustus College}

\begin{center}
{\LARGE Mechanism for the suppression of quantum noise at large scales on
expanding space}

\bigskip

\bigskip

\bigskip

\bigskip

Samuel Colin, Antony Valentini

\textit{Department of Physics and Astronomy,}

\textit{Clemson University, 303 Kinard Laboratory,}

\textit{Clemson, SC 29634-0978, USA.}

\bigskip

\bigskip
\end{center}

\bigskip

\bigskip

\bigskip

\bigskip

\bigskip

\bigskip

We present an exactly-solvable model for the suppression of quantum noise at
large scales on expanding space. The suppression arises naturally in the de
Broglie-Bohm pilot-wave formulation of quantum theory, according to which the
Born probability rule has a dynamical origin. For a scalar field on a
radiation-dominated background we construct the exact solution for the
time-evolving wave functional and study properties of the associated field
trajectories. It is shown that the time evolution of a field mode on expanding
space is mathematically equivalent to that of a standard harmonic oscillator
with a `retarded time' that depends on the wavelength of the mode. In the far
super-Hubble regime the equivalent oscillator evolves over only one Hubble
time, yielding a simple mechanism whereby relaxation to the Born rule can be
suppressed on very large scales. We present numerical simulations illustrating
how the expansion of space can cause a retardation of relaxation in the
super-Hubble regime. Given these results it is natural to expect a suppression
of quantum noise at super-Hubble wavelengths. Such suppression could have
taken place in a pre-inflationary era, resulting in a large-scale power
deficit in the cosmic microwave background.

\bigskip

\bigskip

\bigskip

\bigskip

\bigskip

\bigskip

\bigskip

\bigskip

\bigskip

\bigskip

\bigskip

\bigskip

\bigskip

\bigskip

\bigskip

\bigskip

\bigskip

\bigskip

\bigskip

\section{Introduction}

According to inflationary cosmology \cite{LL00,Muk05,W08,PU09}, the
temperature anisotropy that is observed in the cosmic microwave background
(CMB) was ultimately seeded by quantum fluctuations at very early times.
During inflation the universe undergoes a period of exponential expansion with
a scale factor $a(t)\propto e^{Ht}$ (with $H\approx\mathrm{const}.$). The
expansion is driven by the energy density of an approximately-homogenous
scalar field, whose spatially homogeneous and inhomogeneous parts we
respectively denote $\phi_{0}$ and $\phi$. The anisotropy in the CMB was
generated by primordial curvature perturbations $\mathcal{R}_{\mathbf{k}}$
that in turn were generated by quantum fluctuations of $\phi$ in the
Bunch-Davies vacuum. Measurements of the CMB spectrum may therefore be used to
probe the early quantum vacuum. For this reason inflation has long been
regarded as a testing ground for high-energy physics -- for example to probe
possible high-frequency corrections to the inflationary vacuum state
\cite{BM01,MB01,N01,NP01,KG01,K01,KN01,E01,L02,D02}.

Because the primordial perturbations have a quantum origin, inflation may
equally be used to test quantum theory itself (at very short distances and at
very early times). Several authors have discussed how inflationary CMB
predictions would be affected by a hypothetical dynamical collapse of the wave
function in the early universe (introduced in order to solve the quantum
measurement problem) \cite{PSS06,MVP12,CPS13}. Another line of enquiry
considers the possibility of `quantum nonequilibrium' in the inflationary
vacuum \cite{AV07,AV08a,AV10}, which can arise in the de Broglie-Bohm
pilot-wave formulation of quantum theory \cite{deB28,BV09,B52a,B52b,Holl93}.
Quantum nonequilibrium generates corrections to quantum probabilities without
affecting the quantum state, and therefore changes the spectrum of the vacuum
fluctuations without changing the vacuum wave functional itself. Measurements
of the CMB may then be used to test for the existence of quantum
nonequilibrium at very early times \cite{AV10}.

In pilot-wave theory, a system with configuration $q$ has a wave function
$\psi(q,t)$ obeying the usual Schr\"{o}dinger equation $i\partial\psi/\partial
t=\hat{H}\psi$ (we take $\hbar=1$). In addition, the system has an actual
configuration $q(t)$ evolving in time with a velocity $\dot{q}\equiv dq/dt$
that is determined by $\psi$.\footnote{Note the distinction between a general
point $q$ in configuration space and the actual point $q(t)$ occupied by the
system at time $t$.} For systems with standard Hamiltonians, $\dot{q}$ is
proportional to the gradient $\partial_{q}S$ of the phase $S$ of $\psi$. More
generally, $\dot{q}=j/|\psi|^{2}$ where $j=j\left[  \psi\right]  =j(q,t)$ is
the current associated with the Schr\"{o}dinger equation \cite{SV08}%
.\footnote{At the fundamental level $\psi$ has no \textit{a priori} connection
with probabilities; instead $\psi$ plays the role of a `pilot wave' in
configuration space that guides the motion of an individual system. Because
$\psi$ is a field in configuration space and not an ordinary field in 3-space,
it does not itself carry energy or momentum. For a detailed discussion of the
interpretation of this theory see ref. \cite{AVPwtMw}. Historically, the
theory was first proposed by de Broglie at the 1927 Solvay conference
\cite{BV09}.} The current satisfies a continuity equation%
\begin{equation}
\frac{\partial\left\vert \psi\right\vert ^{2}}{\partial t}+\partial_{q}\cdot
j=0\ .
\end{equation}
For an ensemble of systems with initial wave function $\psi(q,t_{i})$, we may
in principle consider an arbitrary initial distribution $\rho(q,t_{i})$ of
configurations $q(t_{i})$. Because each system has velocity $\dot{q}$, the
time evolution $\rho(q,t)$ of the distribution is determined by the continuity
equation%
\begin{equation}
\frac{\partial\rho}{\partial t}+\partial_{q}\cdot\left(  \rho\dot{q}\right)
=0\ .
\end{equation}
Since $\left\vert \psi\right\vert ^{2}$ obeys the same equation, an initial
distribution $\rho(q,t_{i})=\left\vert \psi(q,t_{i})\right\vert ^{2}$ evolves
into $\rho(q,t)=\left\vert \psi(q,t)\right\vert ^{2}$. This is the state of
`quantum equilibrium', for which the distribution matches the Born probability
rule. But the dynamics also allows us to consider `nonequilibrium'
distributions $\rho(q,t_{i})\neq\left\vert \psi(q,t_{i})\right\vert ^{2}$
\cite{AV91a,AV91b,AV92} -- just as classical mechanics allows us to consider
initial distributions that depart from thermal equilibrium.

It is well known that the empirical predictions of quantum theory follow from
pilot-wave dynamics if it is assumed that the initial ensemble is in quantum
equilibrium, with a distribution $\rho(q,t_{i})=\left\vert \psi(q,t_{i}%
)\right\vert ^{2}$. This was shown fully by Bohm in 1952 \cite{B52a,B52b}. A
key point in the derivation is to apply the dynamics to the measuring
apparatus as well as to the microscopic system. The distribution of apparatus
readings or outcomes then agrees with quantum theory. However, in general the
distribution of outcomes depends on the assumed initial distribution
$\rho(q,t_{i})$ of configurations. For an initial nonequilibrium ensemble,
with $\rho(q,t_{i})\neq\left\vert \psi(q,t_{i})\right\vert ^{2}$, the
distribution of quantum measurement outcomes will generally disagree with the
predictions of quantum theory. Thus, at least in principle, pilot-wave theory
contains a physics that is much wider than quantum physics, with possible
nonequilibrium distributions that violate the usual Born rule
\cite{AV91a,AV91b,AV92,AV96,AV01,AV02,AV07,AV08a,AV09,AV10,PV06}. Such
distributions give rise to new phenomena such as nonlocal signalling
\cite{AV91b} -- acting along an underlying preferred foliation of spacetime
\cite{AV08b} -- and `subquantum' measurements that violate the uncertainty
principle \cite{AV02,PV06}. Quantum physics is then seen as a special
equilibrium case of a much wider nonequilibrium physics.

In pilot-wave theory, the equilibrium state $\rho=\left\vert \psi\right\vert
^{2}$ arises from a process of relaxation that is analogous to classical
thermal relaxation. The $H$-function%
\begin{equation}
H=\int dq\ \rho\ln(\rho/\left\vert \psi\right\vert ^{2})\label{H}%
\end{equation}
(minus the relative entropy of $\rho$ with respect to $\left\vert
\psi\right\vert ^{2}$) quantifies the difference between $\rho$ and
$\left\vert \psi\right\vert ^{2}$. It obeys a coarse-graining $H$-theorem
analogous to the classical one, where the minimum $H=0$ corresponds to
equilibrium \cite{AV91a,AV92,AV01}. For initial wave functions that are
superpositions of different energy eigenfunctions, extensive numerical
evidence shows that initial nonequilibrium distributions $\rho$ rapidly
approach $|\psi|^{2}$ on a coarse-grained level (assuming that the initial
state has no fine-grained micro-structure)
\cite{AV92,AV01,VW05,EC06,TRV12,SC12}, with an approximately exponential decay
of the coarse-grained $H$-function \cite{VW05,TRV12}. All the systems that we
have experimental access to have had a long and violent astrophysical history.
Therefore today we would expect to see quantum equilibrium for these systems
(such as atoms in the laboratory). And indeed experiment has confirmed the
Born rule in a wide range of conditions.

On the other hand it is conceivable that quantum nonequilibrium existed in the
early universe, at very early times before relaxation took place
\cite{AV91a,AV91b,AV92,AV96,AV01,AV07,AV08a,AV09,AV10}. This is certainly a
possibility, in the sense that pilot-wave theory would allow it. It is also
(arguably) to be expected, since an equilibrium state today will naturally
have arisen by a process of relaxation from an earlier nonequilibrium state
(as in ordinary statistical mechanics). Further motivations may be given for
the hypothesis of quantum nonequilibrium in the remote past, including certain
otherwise-puzzling `conspiratorial' or `finely-tuned' features of quantum
theory, and a possible solution to the early homogeneity problem (which
afflicts even some models of inflation) \cite{AV91a,AV91b,AV92,AV96,AV01,AV10}.

Quantum nonequilibrium during the inflationary phase could certainly leave an
imprint today in the CMB. It was shown in ref. \cite{AV10} that, if the
inflaton field $\phi$ is in a state of quantum nonequilibrium at the onset of
inflation, then the power spectrum for primordial curvature perturbations
$\mathcal{R}_{\mathbf{k}}$ will be given by $\mathcal{P}_{\mathcal{R}%
}(k)=\mathcal{P}_{\mathcal{R}}^{\mathrm{QT}}(k)\xi(k)$, where $\mathcal{P}%
_{\mathcal{R}}^{\mathrm{QT}}(k)$ is the usual quantum-theoretical prediction
and $\xi(k)$ is a `nonequilibrium function' that is equal to the ratio of the
nonequilibrium and quantum variances for the Fourier components $\phi
_{\mathbf{k}}$. (It was shown that this ratio is preserved in time during the
inflationary expansion itself.) Measurements of the angular power spectrum
$C_{l}$ for the CMB may then be used to set empirical bounds on $\xi(k)$ --
that is, to set limits on corrections to the Born rule during inflation. The
hypothesis of quantum nonequilibrium at or close to the big bang can therefore
be tested using inflationary cosmology.

A more ambitious task is to predict some features of the function $\xi(k)$.
One possible strategy is to consider a pre-inflationary era and to derive
constraints on residual nonequilibrium from that time. It was suggested in
ref. \cite{AV07} that relaxation could be suppressed for super-Hubble field
modes in a radiation-dominated universe, opening up the possibility that in
some circumstances nonequilibrium would survive until later times. In refs.
\cite{AV08a,AVbook} such suppression is shown to occur by means of an upper
bound on the mean displacement of trajectories in configuration space,
resulting in a `freezing inequality' that implies relaxation suppression for
super-Hubble modes (when the inequality is satisfied). However, the inequality
depends on the unknown time evolution of the quantum state and is difficult to
evaluate. On this basis it was suggested in refs. \cite{AV07,AV08a,AV10} that
-- in a cosmology with a radiation-dominated pre-inflationary phase -- there
would exist a large-scale power deficit in the CMB, above some comoving
wavelength $\lambda_{\mathrm{c}}$ that remained to be estimated. For several
years the existence of an infra-red power deficit in the \textit{WMAP} data
was controversial \cite{B11}, but such a deficit has recently been confirmed
in the \textit{Planck} data \cite{PlanckXV}. The statistical significance is
not high: the deficit might be a mere fluctuation. Even so, it is worth
exploring physical models that predict such a deficit in order to better
assess its nature and significance. Therefore we return to this theme here.

In this paper we present an exactly-solvable model of the suppression of
relaxation for super-Hubble modes on expanding space, resulting in a
suppression of quantum noise at large scales. For a free scalar field in a
radiation-dominated universe we find the exact solution for the time-evolving
wave functional and we demonstrate certain properties of the associated de
Broglie-Bohm trajectories. In particular, we show that the time evolution of a
field mode on expanding space is mathematically equivalent to the time
evolution of a standard harmonic oscillator -- but with real time replaced by
a `retarded time' that depends on the wavelength of the mode. In the far
super-Hubble regime we find that the equivalent oscillator evolves over only
one Hubble time. This result yields a simple mechanism whereby relaxation to
the Born rule can be suppressed at very large scales. We also provide
numerical simulations illustrating how the expansion of space can cause a
suppression (or retardation) of relaxation in the super-Hubble regime. These
exact results broadly confirm the expected relaxation suppression for
super-Hubble modes that was proposed in refs. \cite{AV07,AV08a,AV10}.

In the light of these results, it is natural to expect a suppression of
quantum noise at super-Hubble wavelengths in a radiation-dominated expansion
(if nonequilibrium existed at the beginning of the expansion). As noted, such
suppression could have taken place in a pre-inflationary era, resulting in a
power deficit in the inflationary spectrum above some large wavelength
$\lambda_{\mathrm{c}}$ \cite{AV10}. Here we shall provide a simple estimate of
the cutoff $\lambda_{\mathrm{c}}$, which is found to depend essentially on the
number $N$ of inflationary e-folds and on the reheating temperature
$T_{\mathrm{end}}$ at the end of inflation. We find that the allowed parameter
space for $N$ and $T_{\mathrm{end}}$ is consistent with a cutoff
$\lambda_{\mathrm{c}}$ corresponding to the scale of the power deficit
observed in the CMB by the \textit{Planck} satellite \cite{PlanckXV}. It is
therefore conceivable that the observed deficit is caused by the mechanism
discussed in this paper. It is also quite possible that in the real universe
our $\lambda_{\mathrm{c}}$ is so large as to yield a negligible effect on the
CMB -- for example, if the number $N$ of e-folds is very large. This remains
to be seen. Here we are mostly concerned with demonstrating a general
mechanism for quantum noise suppression at large scales. The detailed
application of this mechanism to specific cosmological models, and an
evaluation of the significance of the results compared with rival models, is
left for future work.

It should be emphasised that a suppression of power at large scales arises
quite naturally in the de Broglie-Bohm formulation of quantum theory,
according to which quantum noise has a dynamical origin. The dynamics itself
generates a rapid relaxation in the sub-Hubble regime and a suppression of
relaxation in the super-Hubble regime. The value of the comoving lengthscale
$\lambda_{\mathrm{c}}$ above which such suppression occurs will, however,
depend on the cosmological model.

Finally, we note that a cosmology with a radiation-dominated pre-inflationary
phase has been considered by some authors \cite{VF82,L82,S82,PK07,WN08}. (For
a discussion of motivations for such a cosmology, see ref. \cite{PK07}.)
Working in terms of standard quantum theory, a pre-inflationary era can yield
corrections to the inflationary vacuum with a resulting power deficit at large
scales \cite{PK07,WN08}. To distinguish the latter effect from that studied
here would require detailed predictions for the nonequilibrium function
$\xi(k)$ (see Section 8).

In Section 2 we present the pilot-wave dynamics of a scalar field on expanding
space. In Section 3 we find the exact solution for the wave function of a
single mode in the case of a radiation-dominated expansion. In Section 4 we
discuss the associated de Broglie-Bohm velocity field and we demonstrate that
the dynamics is equivalent to that of a standard harmonic oscillator with a
retarded time. This result is used in Section 5 to show that quantum
nonequilibrium can be frozen in the far super-Hubble regime. In Section 6 we
present numerical simulations that illustrate the suppression of quantum noise
at super-Hubble wavelengths. In Section 7 we outline a possible application of
this mechanism to cosmology. We briefly review how quantum nonequilibrium in
the inflationary vacuum can cause a large-scale power deficit in the CMB, and
we discuss how such nonequilibrium could arise from a super-Hubble suppression
of relaxation during a pre-inflationary phase. Our conclusions and suggestions
for future work are given in Section 8.

\section{Pilot-wave dynamics of a scalar field on expanding space}

A free, minimally-coupled, and massless scalar field $\phi$ on a curved
spacetime with 4-metric $g_{\mu\nu}$ has a classical Lagrangian density%
\begin{equation}
\mathcal{L}=\frac{1}{2}\sqrt{-g}g^{\mu\nu}\partial_{\mu}\phi\partial_{\nu}%
\phi\ .
\end{equation}
We shall work on an expanding flat space with line element%
\begin{equation}
d\tau^{2}=dt^{2}-a^{2}d\mathbf{x}^{2}\ ,
\end{equation}
where $a=a(t)$ is the scale factor and we take $c=1$. We then have%
\begin{equation}
\mathcal{L}=\tfrac{1}{2}a^{3}\dot{\phi}^{2}-\tfrac{1}{2}a(\mathbf{\nabla}%
\phi)^{2}\ .
\end{equation}
This implies the classical wave equation%
\begin{equation}
\ddot{\phi}+3\frac{\dot{a}}{a}\dot{\phi}-\frac{1}{a^{2}}\nabla^{2}\phi=0\ .
\label{we}%
\end{equation}

It is convenient to work in Fourier space, with components%
\[
\phi_{\mathbf{k}}(t)=\frac{1}{(2\pi)^{3/2}}\int d^{3}\mathbf{x}\;\phi
(\mathbf{x},t)e^{-i\mathbf{k}\cdot\mathbf{x}}\ .
\]
It is usual to take $a_{0}=1$ today, at time $t_{0}$. Physical wavelengths are
then given by $\lambda_{\mathrm{phys}}=a(t)\lambda$, where $\lambda=2\pi/k$ is
the proper wavelength today and $k=|\mathbf{k}|$ is the corresponding wave number.

We may write $\phi_{\mathbf{k}}$ in terms of its real and imaginary parts,%
\[
\phi_{\mathbf{k}}=\frac{\sqrt{V}}{(2\pi)^{3/2}}\left(  q_{\mathbf{k}%
1}+iq_{\mathbf{k}2}\right)  \ ,
\]
where $V$ is a box normalisation volume. The real variables $q_{\mathbf{k}r}$
($r=1$, $2$) are subject to the constraint $q_{\mathbf{k}1}=q_{-\mathbf{k}1}$,
$q_{\mathbf{k}2}=-q_{-\mathbf{k}2}$ (since $\phi$ is real). In terms of these
variables the Lagrangian $L=\int d^{3}\mathbf{x}\;\mathcal{L}$ reads%
\[
L=\sum_{\mathbf{k}r}\frac{1}{2}\left(  a^{3}\dot{q}_{\mathbf{k}r}^{2}%
-ak^{2}q_{\mathbf{k}r}^{2}\right)  \ .
\]
We then have canonical momenta $\pi_{\mathbf{k}r}\equiv\partial L/\partial
\dot{q}_{\mathbf{k}r}=a^{3}\dot{q}_{\mathbf{k}r}$ and the Hamiltonian becomes%
\[
H=\sum_{\mathbf{k}r}\left(  \frac{1}{2a^{3}}\pi_{\mathbf{k}r}^{2}+\frac{1}%
{2}ak^{2}q_{\mathbf{k}r}^{2}\right)  \ .
\]

This system is readily quantised. The Schr\"{o}dinger equation for $\Psi
=\Psi\lbrack q_{\mathbf{k}r},t]$ reads%
\begin{equation}
i\frac{\partial\Psi}{\partial t}=\sum_{\mathbf{k}r}\left(  -\frac{1}{2a^{3}%
}\frac{\partial^{2}}{\partial q_{\mathbf{k}r}^{2}}+\frac{1}{2}ak^{2}%
q_{\mathbf{k}r}^{2}\right)  \Psi\ . \label{Sch2'}%
\end{equation}
This implies the continuity equation%
\[
\frac{\partial\left\vert \Psi\right\vert ^{2}}{\partial t}+\sum_{\mathbf{k}%
r}\frac{\partial}{\partial q_{\mathbf{k}r}}\left(  \left\vert \Psi\right\vert
^{2}\frac{1}{a^{3}}\frac{\partial S}{\partial q_{\mathbf{k}r}}\right)  =0\ ,
\]
from which we may identify the de Broglie velocities%
\begin{equation}
\frac{dq_{\mathbf{k}r}}{dt}=\frac{1}{a^{3}}\frac{\partial S}{\partial
q_{\mathbf{k}r}}=\frac{1}{a^{3}}\operatorname{Im}\frac{1}{\Psi}\frac
{\partial\Psi}{\partial q_{\mathbf{k}r}} \label{deB2}%
\end{equation}
(with $\Psi=\left\vert \Psi\right\vert e^{iS}$). We may now consider a
theoretical ensemble of fields with the same wave functional $\Psi$. The time
evolution of each field is determined by (\ref{deB2}). The time evolution of
an arbitrary distribution $P[q_{\mathbf{k}r},t]$ of fields will therefore be
determined by%
\begin{equation}
\frac{\partial P}{\partial t}+\sum_{\mathbf{k}r}\frac{\partial}{\partial
q_{\mathbf{k}r}}\left(  P\frac{1}{a^{3}}\frac{\partial S}{\partial
q_{\mathbf{k}r}}\right)  =0\ . \label{ContP1}%
\end{equation}
As usual in pilot-wave theory, if $P[q_{\mathbf{k}r},t_{i}]=\left\vert
\Psi\lbrack q_{\mathbf{k}r},t_{i}]\right\vert ^{2}$ then $P[q_{\mathbf{k}%
r},t]=\left\vert \Psi\lbrack q_{\mathbf{k}r},t]\right\vert ^{2}$ for all $t$.
Whereas if $P[q_{\mathbf{k}r},t_{i}]\neq\left\vert \Psi\lbrack q_{\mathbf{k}%
r},t_{i}]\right\vert ^{2}$ then for as long as $P$ remains in nonequilibrium
the statistics will generally differ from those predicted by the Born rule.

This pilot-wave model has been applied to inflationary cosmology
\cite{AV07,AV08a,AV10}. We have assumed that there is a preferred foliation of
spacetime with time function $t$. (Note that spatial homogeneity is not
required. A similar construction may be given in any globally-hyperbolic
spacetime by choosing a preferred foliation \cite{AV04,AV08b,AVbook}.)

Let us now focus on the case of a decoupled (that is, unentangled) mode
$\mathbf{k}$. If $\Psi$ takes the form $\Psi=\psi_{\mathbf{k}}(q_{\mathbf{k}%
1},q_{\mathbf{k}2},t)\varkappa$, where $\varkappa$ depends only on degrees of
freedom for modes $\mathbf{k}^{\prime}\neq\mathbf{k}$, we may write an
independent dynamics for the mode. Dropping the index $\mathbf{k}$ hereafter,
and introducing the time-dependent quantities%
\begin{equation}
m=a^{3}\ ,\ \ \ \omega=k/a\ ,
\end{equation}
it follows from (\ref{Sch2'}) that the wave function $\psi=\psi(q_{1}%
,q_{2},t)$ satisfies a Schr\"{o}dinger equation%
\begin{equation}
i\frac{\partial\psi}{\partial t}=\sum_{r=1,\ 2}\left(  -\frac{1}{2m}%
\partial_{r}^{2}+\frac{1}{2}m\omega^{2}q_{r}^{2}\right)  \psi\ , \label{S2D}%
\end{equation}
while from (\ref{deB2}) it follows that de Broglie's equation of motion for
the configuration $(q_{1},q_{2})$ reads%
\begin{equation}
\dot{q}_{r}=\frac{1}{m}\operatorname{Im}\frac{\partial_{r}\psi}{\psi}
\label{deB2D}%
\end{equation}
(with $\partial_{r}\equiv\partial/\partial q_{r}$). The marginal distribution
$\rho=\rho(q_{1},q_{2},t)$ for the mode will then evolve according to%
\begin{equation}
\frac{\partial\rho}{\partial t}+\sum_{r=1,\ 2}\partial_{r}\left(  \rho\frac
{1}{m}\operatorname{Im}\frac{\partial_{r}\psi}{\psi}\right)  =0\ . \label{C2D}%
\end{equation}

Equations (\ref{S2D}), (\ref{deB2D}) and (\ref{C2D}) are formally the same as
those of pilot-wave dynamics for a nonrelativistic particle with a
time-dependent mass $m=a^{3}$ and moving (in the $q_{1}-q_{2}$ plane) in a
harmonic oscillator potential with time-dependent angular frequency
$\omega=k/a$. Thus, for a decoupled field mode, we may discuss relaxation (and
its suppression) in terms of relaxation for a nonrelativistic two-dimensional
harmonic oscillator with a time-dependent mass and frequency
\cite{AV07,AV08a,AVbook}.

In a case where $m$ and $\omega$ are constant, it is already known that the
equations (\ref{S2D})--(\ref{C2D}) generate an efficient relaxation to
equilibrium. How will relaxation be affected by a time-dependent scale factor
$a(t)$? The answer depends on how the physical wavelength $\lambda
_{\mathrm{phys}}$ compares with the Hubble radius $H^{-1}$.

In the short-wavelength limit we should recover the equations for a decoupled
mode $\mathbf{k}$ on Minkowski spacetime -- since, roughly speaking, the
timescale $\Delta t\propto\lambda_{\mathrm{phys}}$ over which $\psi$ evolves
will be much smaller than the expansion timescale $H^{-1}\equiv a/\dot{a}$
\cite{AV07}. More precisely, the short-wavelength limit may be defined by
$\lambda_{\mathrm{phys}}<<\Delta n\cdot H^{-1}$, where $n=n_{1}+n_{2}$ is the
sum of the occupation numbers (for the modes $r=1,2$) and $\Delta n$ is the
quantum spread thereof. If we consider an evolution over timescales $\Delta
t\equiv1/\Delta E<<H^{-1}$ (for which $a$ is approximately constant and where
$\Delta E$ is the quantum energy spread), then the equations reduce to those
for a decoupled mode on Minkowski spacetime -- or, to those for a
two-dimensional oscillator with constant mass $m$ and constant angular
frequency $\omega$ \cite{AV08a,AVbook}. Thus, in the far sub-Hubble regime we
may deduce that, if the mode $\mathbf{k}$ is in a superposition of many
different states of definite occupation number, then an initial nonequilibrium
distribution $\rho\neq\left\vert \psi\right\vert ^{2}$ will rapidly relax to
equilibrium (on a coarse-grained level) -- just as occurs for nonrelativistic
particles moving in two dimensions \cite{AV92,AV01,VW05,EC06,TRV12,SC12}.

In contrast, in the long-wavelength limit -- which may be defined by
$\lambda_{\mathrm{phys}}>>\Delta n\cdot H^{-1}$ -- we expect that the wave
function $\psi$ will be approximately static (or `frozen') over timescales
$\sim H^{-1}$. A similar `freezing' over timescales $\sim H^{-1}$ is then
expected both for the trajectories $(q_{1}(t),q_{2}(t))$ and for arbitrary
nonequilibrium distributions $\rho\neq\left\vert \psi\right\vert ^{2}$
\cite{AV07,AV08a}. This is of course reminiscent of the freezing of
super-Hubble modes in the theory of cosmological perturbations
\cite{Pad93,LL00,PU09}. In both cases, the freezing occurs for dynamical reasons.

This simple reasoning suggests that relaxation to quantum equilibrium will
take place as usual in the far sub-Hubble regime but can be suppressed in the
far super-Hubble regime \cite{AV07}. This expectation is supported by a
general upper bound on the mean displacement of trajectories in configuration
space, which implies a suppression of relaxation provided a certain `freezing
inequality' is satisfied \cite{AV08a,AVbook}. The inequality is difficult to
evaluate since it depends on the time evolution of the quantum state; even so,
the inequality can be satisfied only for super-Hubble modes.

Here we construct an exactly-solvable model of relaxation suppression in the
super-Hubble regime. As we shall see, the results broadly confirm the general
expectations \cite{AV07,AV08a,AV10,AVbook}.

\section{Exact solution for the wave function}

We need to solve the Schr\"{o}dinger equation (\ref{S2D}) for the wave
function $\psi=\psi(q_{1},q_{2},t)$. We may write the Hamiltonian as $\hat
{H}=\hat{H}_{1}+\hat{H}_{2}$ where%

\begin{equation}
\hat{H}_{r}=-\frac{1}{2m}\partial_{r}^{2}+\frac{1}{2}m\omega^{2}q_{r}^{2}\ .
\label{ham_1dshoes}%
\end{equation}
The Hamiltonian is of course time dependent, $\hat{H}=\hat{H}(t)$. We shall be
interested in a radiation-dominated expansion, over a time interval
$(t_{i},t_{f})$, with scale factor $a=a_{i}(t/t_{i})^{1/2}$.

To solve this problem we may expand the initial wave function (at time
$t=t_{i}$) in terms of the instantaneous eigenstates $\Phi_{n_{1}}(q_{1}%
)\Phi_{n_{2}}(q_{2})$ of the initial Hamiltonian $\hat{H}(t_{i})$:%
\begin{equation}
\psi(q_{1},q_{2},t_{i})=\sum_{n_{1}n_{2}}c_{n_{1}n_{2}}(t_{i})\Phi_{n_{1}%
}(q_{1})\Phi_{n_{2}}(q_{2})~, \label{psiti}%
\end{equation}
where $\Phi_{n_{r}}(q_{r})$ is the $n_{r}$th eigenstate of the initial
one-dimensional Hamiltonian $\hat{H}_{r}(t_{i})$. If we know how the initial
wave function $\psi_{n_{r}}(q_{r},t_{i})=\Phi_{n_{r}}(q_{r})$ evolves under
the one-dimensional Schr\"{o}dinger equation%
\begin{equation}
i\frac{\partial\psi_{n_{r}}(q_{r},t)}{\partial t}=\hat{H}_{r}(t)\psi_{n_{r}%
}(q_{r},t) \label{se_1dshoes}%
\end{equation}
then we will have the solution to the full two-dimensional problem. The exact
solution for the wave function may then be written as%
\begin{equation}
\psi(q_{1},q_{2},t)=\sum_{n_{1}n_{2}}c_{n_{1}n_{2}}(t_{i})\psi_{n_{1}}%
(q_{1},t)\psi_{n_{2}}(q_{2},t)~. \label{exact psi}%
\end{equation}

The problem is therefore reduced to solving (\ref{se_1dshoes}) for all of the
wave functions $\psi_{n_{r}}=\psi_{n_{r}}(q_{r},t)$ with the initial
conditions $\psi_{n_{r}}(q_{r},t_{i})=\Phi_{n_{r}}(q_{r})$. A partial answer
to this problem can be found in ref. \cite{Ji95} (building on the early work
of Lewis and Riesenfeld \cite{L67,LR69}), where it is shown that the required
wave functions take the form%
\begin{align}
\psi_{n_{r}}(q_{r},t)  &  =\frac{1}{\sqrt{2^{n_{r}}n_{r}!}}\left(
\frac{\omega_{i}}{\pi g_{-}(t)}\right)  ^{\frac{1}{4}}\exp\left(
-i\frac{g_{0}(t)}{2g_{-}(t)}q_{r}^{2}\right)  .\exp\left(  -i(n_{r}+\frac
{1}{2})\int_{t_{i}}^{t}dt^{\prime}\frac{\omega_{i}}{m(t^{\prime}%
)g_{-}(t^{\prime})}\right) \nonumber\\
&  \times\exp\left(  -\frac{\omega_{i}}{2g_{-}(t)}q_{r}^{2}\right)
.\mathcal{H}_{n_{r}}\left(  \sqrt{\frac{\omega_{i}}{g_{-}(t)}}q_{r}\right)
\ . \label{psint}%
\end{align}
Here $\omega_{i}=\omega(t_{i})$, the $\mathcal{H}_{n}$ are Hermite
polynomials, and the functions $g_{-}(t),\ g_{0}(t)$ and $g_{+}(t)$ satisfy
the ordinary differential equations (valid for general $a(t)$)%
\begin{align}
\dot{g}_{-}  &  =-2\frac{g_{0}}{m}\label{eqdiff1}\\
\dot{g}_{0}  &  =m\omega^{2}g_{-}-\frac{g_{+}}{m}\label{eqdiff2}\\
\dot{g}_{+}  &  =2m\omega^{2}g_{0} \label{eqdiff3}%
\end{align}
with the initial conditions%
\begin{equation}
g_{-}(t_{i})=\frac{1}{m_{i}}\text{~},\text{\ \ \ }g_{0}(t_{i})=0\text{~}%
,\text{\ \ \ }g_{+}(t_{i})=m_{i}\omega_{i}^{2} \label{initcond}%
\end{equation}
(where $m_{i}=m(t_{i})$). According to the analysis of ref. \cite{Ji95}, the
most general solution for $g_{-}(t)$ takes the form
\begin{equation}
g_{-}=c_{1}f_{1}^{2}+c_{2}f_{1}f_{2}+c_{3}f_{2}^{2}~, \label{gminus0}%
\end{equation}
where $f_{1}(t)$ and $f_{2}(t)$ are two independent solutions of the classical
equation of motion
\begin{equation}
\ddot{f}+\frac{\dot{m}}{m}\dot{f}+\omega^{2}f=0~. \label{classical_shoes}%
\end{equation}
If two independent solutions of (\ref{classical_shoes}) can be found, we will
have an expression for $g_{-}$ involving the three constants $c_{1},\ c_{2}$
and $c_{3}$. The functions $g_{0}$, $g_{+}$ can then be determined from
$g_{-}$ by means of (\ref{eqdiff1}) and (\ref{eqdiff2}). Finally, the
constants $c_{1}$, $c_{2}$ and $c_{3}$ are fixed by the initial conditions
(\ref{initcond}).

Equation (\ref{classical_shoes}) is of course the well-known equation for
modes $\phi(\mathbf{x},t)\propto f_{1}(t)e^{i\mathbf{k}\cdot\mathbf{x}}$ of
the wave equation (\ref{we}). For any power law $a\propto t^{p}$ it has
solutions that are Bessel functions \cite{PU09}. For definiteness, we shall
restrict ourselves to the case $a\propto t^{1/2}$.

\subsection{Solution for a radiation-dominated expansion}

We require the solutions (\ref{psint}) for a radiation-dominated expansion
$a=a_{i}(t/t_{i})^{1/2}$. To this end we must first obtain two independent
solutions of (\ref{classical_shoes}). With $m=a^{3}=a_{i}^{3}(t/t_{i})^{3/2}$
and $\omega=k/a=(k/a_{i})(t_{i}/t)^{1/2}$, equation (\ref{classical_shoes})
becomes
\begin{equation}
\ddot{f}(t)+\frac{3}{2t}\dot{f}(t)+\frac{\varepsilon}{t}f(t)=0~,
\label{classical_shoes2}%
\end{equation}
where it is useful to define the parameter%
\begin{equation}
\varepsilon\equiv\left(  \frac{t_{i}}{a_{i}^{2}}\right)  k^{2} \label{epsilon}%
\end{equation}
(so that $\omega^{2}=\varepsilon/t$). The solutions to (\ref{classical_shoes2}%
) are%
\begin{equation}
f_{1}=\frac{1}{\sqrt{t}}\cos2(\sqrt{\varepsilon t}-\sqrt{\varepsilon t_{i}%
}),\ \text{ }f_{2}=\frac{1}{\sqrt{t}}\sin2(\sqrt{\varepsilon t}-\sqrt
{\varepsilon t_{i}})~.
\end{equation}
From (\ref{gminus0}) it then follows that%
\begin{align}
g_{-}  &  =\frac{c_{1}}{t}\cos^{2}2(\sqrt{\varepsilon t}-\sqrt{\varepsilon
t_{i}})+\nonumber\\
&  \frac{c_{2}}{t}\sin2(\sqrt{\varepsilon t}-\sqrt{\varepsilon t_{i}}%
).\cos2(\sqrt{\varepsilon t}-\sqrt{\varepsilon t_{i}})+\frac{c_{3}}{t}\sin
^{2}2(\sqrt{\varepsilon t}-\sqrt{\varepsilon t_{i}})~.
\end{align}
This can be rewritten as%
\begin{equation}
g_{-}=\frac{1}{t}\left(  A+B\cos4(\sqrt{\varepsilon t}-\sqrt{\varepsilon
t_{i}})+C\sin4(\sqrt{\varepsilon t}-\sqrt{\varepsilon t_{i}})\right)
\label{gminus_exact}%
\end{equation}
where $A$, $B$ and $C$ are three constants that need to be determined.

To fix $A$, $B$ and $C$, we first use (\ref{eqdiff1}) and (\ref{eqdiff2}) to
calculate $g_{0}$, $g_{+}$ from $g_{-}$ and we then impose the initial
conditions (\ref{initcond}). We find that%
\begin{equation}
A=\frac{1+8\varepsilon t_{i}}{8a_{i}^{3}\varepsilon}\ ,\ \ \ B=-\frac
{1}{8a_{i}^{3}\varepsilon}\ ,\ \ \ C=\frac{\sqrt{\varepsilon t_{i}}}%
{2a_{i}^{3}\varepsilon}~. \label{constants}%
\end{equation}
Thus we have%
\begin{equation}
g_{-}(t)=\frac{1}{8a_{i}^{3}\varepsilon t}\left[  (1+8\varepsilon t_{i}%
)-\cos{4(\sqrt{\varepsilon t}-\sqrt{\varepsilon t_{i}})}+4\sqrt{\varepsilon
t_{i}}\sin{4(\sqrt{\varepsilon t}-\sqrt{\varepsilon t_{i}})}\right]  ~.
\label{gminus}%
\end{equation}
We also find that%
\begin{equation}
g_{0}(t)=\frac{1}{16\varepsilon t_{i}\sqrt{t_{i}t}}\left(
\begin{array}
[c]{c}%
(1+8\varepsilon t_{i})-(1+8\sqrt{\varepsilon t_{i}}\sqrt{\varepsilon t}%
)\cos{4(\sqrt{\varepsilon t}-\sqrt{\varepsilon t_{i}})}\\
+2(2\sqrt{\varepsilon t_{i}}-\sqrt{\varepsilon t})\sin{4(\sqrt{\varepsilon
t}-\sqrt{\varepsilon t_{i}})}%
\end{array}
\right) \nonumber
\end{equation}

Finally, to have the complete expression for the wave functions (\ref{psint})
we must evaluate the integral%
\begin{equation}
\Theta(t)\equiv\int_{t_{i}}^{t}dt^{\prime}\frac{\omega_{i}}{m(t^{\prime}%
)g_{-}(t^{\prime})}\ . \label{Theta}%
\end{equation}
Using $m=a^{3}$ and (\ref{gminus_exact}), the integral takes the form%
\begin{equation}
\Theta=\frac{k}{a_{i}}\frac{t_{i}^{3/2}}{a_{i}^{3}}\int_{t_{i}}^{t}%
\frac{dt^{\prime}}{\sqrt{t^{\prime}}}\frac{1}{A+B\cos4(\sqrt{\varepsilon
t^{\prime}}-\sqrt{\varepsilon t_{i}})+C\sin4(\sqrt{\varepsilon t^{\prime}%
}-\sqrt{\varepsilon t_{i}})}~.
\end{equation}
With the change of variables $\phi=4(\sqrt{\varepsilon t^{\prime}}%
-\sqrt{\varepsilon t_{i}})$ we have%
\begin{equation}
\Theta=\frac{kt_{i}^{3/2}}{2a_{i}^{4}\sqrt{\varepsilon}}\int_{0}%
^{4(\sqrt{\varepsilon t}-\sqrt{\varepsilon t_{i}})}d\phi\frac{1}{A+B\cos
\phi+C\sin\phi}~.
\end{equation}
This integral may be evaluated, with a result that depends on the relation
between $A$, $B$ and $C$. From (\ref{constants}) we have $A^{2}>B^{2}+C^{2}$.
In this case one has (ref. \cite{RR}, p.174)%
\begin{equation}
\int d\phi\frac{1}{A+B\cos\phi+C\sin\phi}=\frac{2}{\sqrt{A^{2}-B^{2}-C^{2}}%
}\tan^{-1}{\frac{(A-B)\tan{\frac{\phi}{2}}+C}{\sqrt{A^{2}-B^{2}-C^{2}}}\ .}%
\end{equation}
This result employs the change of variables $t=\tan{\frac{\phi}{2}}$, which is
singular when $\phi$ is an odd multiple of $\pi$. Therefore the domain of
integration must be cut into parts ($[0,\pi)$, $(\pi,3\pi)$, \ldots) and each
time $\phi$ moves from one domain to the next there is an additional
contribution of $\pi$ which must be added to the function $\tan^{-1}$.
Therefore the total result for $\Theta$ is%

\begin{align}
\Theta(t)  &  =\tan^{-1}\left(  {\frac{1+4\varepsilon t_{i}}{4\varepsilon
t_{i}}\tan{(2\sqrt{\varepsilon t}-2\sqrt{\varepsilon t_{i}})}+\frac{1}%
{2\sqrt{\varepsilon t_{i}}}}\right) \nonumber\\
&  +\pi.\text{nint}(\frac{2\sqrt{\varepsilon t}-2\sqrt{\varepsilon t_{i}}}%
{\pi})-\tan^{-1}\left(  \frac{1}{2\sqrt{\varepsilon t_{i}}}\right)  ~.
\label{Thetaeval}%
\end{align}
(where $\text{nint}(x)$ returns the integer nearest to $x$).

\section{Properties of the trajectories}

We have an exact solution (\ref{exact psi}) for the wave function $\psi
(q_{1},q_{2},t)$. In pilot-wave theory the actual configuration $(q_{1}%
(t),q_{2}(t))$ at time $t$ evolves according to de Broglie's equation of
motion (\ref{deB2D}). This yields velocities%
\begin{equation}
\dot{q}_{1}=-\frac{1}{m}\frac{g_{0}}{g_{-}}q_{1}+\frac{1}{m}\sqrt{\frac
{\omega_{i}}{g_{-}}}\operatorname{Im}\left(  \frac{\sum_{n_{1}n_{2}}{\tilde
{c}}_{n_{1}n_{2}}(t_{ret}(t))\mathcal{H}_{n_{1}}^{\prime}(\sqrt{\frac
{\omega_{i}}{g_{-}}}q_{1})\mathcal{H}_{n_{2}}(\sqrt{\frac{\omega_{i}}{g_{-}}%
}q_{2})}{\sum_{m_{1}m_{2}}\tilde{c}_{m_{1}m_{2}}(t_{ret}(t))\mathcal{H}%
_{m_{1}}(\sqrt{\frac{\omega_{i}}{g_{-}}}q_{1})\mathcal{H}_{m_{2}}(\sqrt
{\frac{\omega_{i}}{g_{-}}}q_{2})}\right)  \label{mq1dot}%
\end{equation}
and%
\begin{equation}
\dot{q}_{2}=-\frac{1}{m}\frac{g_{0}}{g_{-}}q_{2}+\frac{1}{m}\sqrt{\frac
{\omega_{i}}{g_{-}}}\operatorname{Im}\left(  \frac{\sum_{n_{1}n_{2}}{\tilde
{c}}_{n_{1}n_{2}}(t_{ret}(t))\mathcal{H}_{n_{1}}(\sqrt{\frac{\omega_{i}}%
{g_{-}}}q_{1})\mathcal{H}_{n_{2}}^{\prime}(\sqrt{\frac{\omega_{i}}{g_{-}}%
}q_{2})}{\sum_{m_{1}m_{2}}\tilde{c}_{m_{1}m_{2}}(t_{ret}(t))\mathcal{H}%
_{m_{1}}(\sqrt{\frac{\omega_{i}}{g_{-}}}q_{1})\mathcal{H}_{m_{2}}(\sqrt
{\frac{\omega_{i}}{g_{-}}}q_{2})}\right)  \label{mq2dot}%
\end{equation}
(where a prime on $\mathcal{H}$ denotes a derivative with respect to the
argument). Here%
\begin{equation}
\tilde{c}_{n_{1}n_{2}}(t)\equiv\frac{{c}_{n_{1}n_{2}}(t_{i})e^{-i(t-t_{i}%
)\omega_{i}(n_{1}+n_{2}+1)}}{\sqrt{2^{n_{1}}{n_{1}}!}\sqrt{2^{n_{2}}{n_{2}}!}}%
\end{equation}
and we have defined the \textit{retarded time}%
\begin{equation}
t_{\mathrm{ret}}(t)\equiv t_{i}+\int_{t_{i}}^{t}\frac{1}{m(t^{\prime}%
)g_{-}(t^{\prime})}dt^{\prime}~. \label{tr}%
\end{equation}
This is related to $\Theta(t)$ by%
\begin{equation}
t_{\mathrm{ret}}(t)=t_{i}+\frac{1}{\omega_{i}}\Theta(t)~.
\end{equation}
(For a plot of the function $t_{\mathrm{ret}}=t_{\mathrm{ret}}(t)$, see Figure 1.)

\subsection{Rescaled variables}

Our wave functions $\psi_{n_{r}}(q_{r},t)$, given by (\ref{psint}), have
time-dependent widths that are proportional to $\sqrt{g_{-}}$. From the
solution (\ref{gminus}) for $g_{-}$ ($\sim1/t$), we see that the widths shrink
with time as $\sim1/\sqrt{t}$ (with an oscillatory factor as well). Because of
this shrinking support it is convenient to use the rescaled variables%
\begin{equation}
q_{r}^{\prime}=\sqrt{\frac{\omega_{i}}{g_{-}(t)}}q_{r}\ .
\end{equation}
Their time evolution is given by%
\begin{equation}
\frac{dq_{r}^{\prime}}{dt}=\sqrt{\frac{\omega_{i}}{g_{-}}}\left(  \dot{q}%
_{r}-\frac{1}{2}\frac{\dot{g}_{-}}{g_{-}}q_{r}\right)  =\sqrt{\frac{\omega
_{i}}{g_{-}}}\left(  \dot{q}_{r}+\frac{1}{m}\frac{g_{0}}{g_{-}}q_{r}\right)
~, \label{vbar}%
\end{equation}
where we have used (\ref{eqdiff1}). From the respective expressions
(\ref{mq1dot}), (\ref{mq2dot}) for $\dot{q}_{1}$, $\dot{q}_{2}$ we then find%
\begin{equation}
\dot{q}_{1}^{\prime}(t)=\frac{1}{m}\frac{\omega_{i}}{g_{-}}\operatorname{Im}%
\left[  \frac{\sum_{n_{1}n_{2}}\tilde{c}_{n_{1}n_{2}}(t_{ret}(t))\mathcal{H}%
_{n_{1}}^{\prime}(q_{1}^{\prime})\mathcal{H}_{n_{2}}(q_{2}^{\prime})}%
{\sum_{m_{1}m_{2}}\tilde{c}_{m_{1}m_{2}}(t_{ret}(t))\mathcal{H}_{m_{1}}%
(q_{1}^{\prime})\mathcal{H}_{m_{2}}(q_{2}^{\prime})}\right]  \label{vbarq1}%
\end{equation}
and%
\begin{equation}
\dot{q}_{2}^{\prime}(t)=\frac{1}{m}\frac{\omega_{i}}{g_{-}}\operatorname{Im}%
\left[  \frac{\sum_{n_{1}n_{2}}\tilde{c}_{n_{1}n_{2}}(t_{ret}(t))\mathcal{H}%
_{n_{1}}(q_{1}^{\prime})\mathcal{H}_{n_{2}}^{\prime}(q_{2}^{\prime})}%
{\sum_{m_{1}m_{2}}\tilde{c}_{m_{1}m_{2}}(t_{ret}(t))\mathcal{H}_{m_{1}}%
(q_{1}^{\prime})\mathcal{H}_{m_{2}}(q_{2}^{\prime})}\right]  ~. \label{vbarq2}%
\end{equation}

\subsection{Equivalence to the standard oscillator at retarded time}

We shall now show that the rescaled trajectory of the system is identical to a
rescaled trajectory generated by a \textit{standard} harmonic oscillator --
with coordinates $Q_{r}$, constant mass $m_{i}$ and constant frequency
$\omega_{i}$, and with the same initial wave function (\ref{psiti}) at
$t=t_{i}$ -- but now with time running from $t_{i}$ up to the retarded time
$t_{\mathrm{ret}}(t)$ (instead of from $t_{i}$ to $t$). We also demonstrate a
correspondence between the equilibrium states for the two systems.

For our standard harmonic oscillator we have a wave function $\psi
_{\mathrm{SHO}}=\psi_{\mathrm{SHO}}(Q_{1},Q_{2},t)$ and a Schr\"{o}dinger
equation%
\begin{equation}
i\frac{\partial\psi_{\mathrm{SHO}}}{\partial t}=-\frac{1}{2m_{i}}\left(
\frac{\partial^{2}}{\partial Q_{1}^{2}}+\frac{\partial^{2}}{\partial Q_{2}%
^{2}}\right)  \psi_{\mathrm{SHO}}+\frac{1}{2}m_{i}\omega_{i}^{2}(Q_{1}%
^{2}+Q_{2}^{2})\psi_{\mathrm{SHO}}\ .
\end{equation}
For $Q_{r}$ the de Broglie velocity field is given by%
\begin{equation}
\dot{Q}_{r}=\frac{1}{m_{i}}\operatorname{Im}\frac{1}{\psi_{\mathrm{SHO}}}%
\frac{\partial\psi_{\mathrm{SHO}}}{\partial Q_{r}}\ . \label{deBQ}%
\end{equation}

With an initial wave function%
\[
\psi_{\mathrm{SHO}}(Q_{1},Q_{2},t_{i})=\sum_{n_{1}n_{2}}c_{n_{1}n_{2}}%
(t_{i})\Phi_{n_{1}}(Q_{1})\Phi_{n_{2}}(Q_{2})
\]
(identical to (\ref{psiti}) with the coordinates $q_{1},\ q_{2}$ replaced by
$Q_{1},\ Q_{2}$), we have the solution%
\begin{equation}
\psi_{\mathrm{SHO}}(Q_{1},Q_{2},t)=\sum_{n_{1}n_{2}}c_{n_{1}n_{2}}(t_{i}%
)\psi_{\mathrm{SHO\,}n_{1}}(Q_{1},t)\psi_{\mathrm{SHO\,}n_{2}}(Q_{2},t)\ ,
\label{exact psi SHO}%
\end{equation}
where now instead of (\ref{psint}) the functions $\psi_{\mathrm{SHO\,}n_{r}%
}(Q_{r},t)$ take the simple form%
\begin{align}
\psi_{\mathrm{SHO}\,n_{r}}(Q_{r},t)  &  =\frac{1}{\sqrt{2^{n_{r}}n_{r}!}%
}\left(  \frac{m_{i}\omega_{i}}{\pi}\right)  ^{\frac{1}{4}}\exp\left(
-i(n_{r}+\frac{1}{2})\omega_{i}(t-t_{i})\right) \nonumber\\
&  \times\exp\left(  -\frac{m_{i}\omega_{i}}{2}Q_{r}^{2}\right)
.\mathcal{H}_{n_{r}}\left(  \sqrt{m_{i}\omega_{i}}Q_{r}\right)  \ .
\end{align}
Introducing the rescaled variable $Q_{r}^{\prime}=\sqrt{m_{i}\omega_{i}}Q_{r}$
(which has the same rescaling as $q_{r}^{\prime}$ at $t=t_{i}$), it follows
from (\ref{deBQ}) that the velocities $\dot{Q}_{r}^{\prime}=\sqrt{m_{i}%
\omega_{i}}\dot{Q}_{r}$ are given by%
\begin{equation}
\dot{Q}_{1}^{\prime}(t)=\omega_{i}\operatorname{Im}\left[  \frac{\sum
_{n_{1}n_{2}}\tilde{c}_{n_{1}n_{2}}(t)\mathcal{H}_{n_{1}}^{\prime}%
(Q_{1}^{\prime})\mathcal{H}_{n_{2}}(Q_{2}^{\prime})}{\sum_{m_{1}m_{2}}%
\tilde{c}_{m_{1}m_{2}}(t)\mathcal{H}_{m_{1}}(Q_{1}^{\prime})\mathcal{H}%
_{m_{2}}(Q_{2}^{\prime})}\right]  \label{vbarQ1}%
\end{equation}
and%
\begin{equation}
\dot{Q}_{2}^{\prime}(t)=\omega_{i}\operatorname{Im}\left[  \frac{\sum
_{n_{1}n_{2}}\tilde{c}_{n_{1}n_{2}}(t)\mathcal{H}_{n_{1}}(Q_{1}^{\prime
})\mathcal{H}_{n_{2}}^{\prime}(Q_{2}^{\prime})}{\sum_{m_{1}m_{2}}\tilde
{c}_{m_{1}m_{2}}(t)\mathcal{H}_{m_{1}}(Q_{1}^{\prime})\mathcal{H}_{m_{2}%
}(Q_{2}^{\prime})}\right]  ~. \label{vbarQ2}%
\end{equation}

Let us now compare the velocities for $q_{r}^{\prime}$ and $Q_{r}^{\prime}$.
We have, from (\ref{vbarq1}), (\ref{vbarq2}) and (\ref{vbarQ1}),
(\ref{vbarQ2}), the simple relationship%
\begin{equation}
\dot{q}_{r}^{\prime}(t)|_{q_{1}^{\prime}=a,\ q_{2}^{\prime}=b}=\frac
{1}{m(t)g_{-}(t)}\dot{Q}_{r}^{\prime}(t_{\mathrm{ret}}(t))|_{Q_{1}^{\prime
}=a,\ Q_{2}^{\prime}=b} \label{vVret}%
\end{equation}
or%
\begin{equation}
\dot{q}_{r}^{\prime}(t)|_{q_{1}^{\prime}=a,\ q_{2}^{\prime}=b}dt=\dot{Q}%
_{r}^{\prime}(t_{\mathrm{ret}}(t))|_{Q_{1}^{\prime}=a,\ Q_{2}^{\prime}%
=b}dt_{\mathrm{ret}}\ , \label{vVret2}%
\end{equation}
where we have used%
\begin{equation}
dt_{\mathrm{ret}}=\frac{1}{m(t)g_{-}(t)}dt\ . \label{dtret}%
\end{equation}
Here $t_{\mathrm{ret}}=t_{\mathrm{ret}}(t)$ is the retarded time (\ref{tr})
`corresponding to' time $t$, and each side of (\ref{vVret}) or (\ref{vVret2})
is evaluated at the same point $(a,b)$ in the respective configuration space.

The displacement of $q_{r}^{\prime}$ from $t_{i}$ up to time $t$ is given by%
\[
\delta q_{r}^{\prime}(t,t_{i})=\int_{t_{i}}^{t}\dot{q}_{r}^{\prime}(t^{\prime
})dt^{\prime}~,
\]
while the displacement of $Q_{r}^{\prime}$ from $t_{i}$ up to the
corresponding retarded time $t_{\mathrm{ret}}(t)$ is given by%
\[
\delta Q_{r}^{\prime}(t_{\mathrm{ret}},t_{i})=\int_{t_{i}}^{t_{\mathrm{ret}}%
}\dot{Q}_{r}^{\prime}(t^{\prime})dt^{\prime}~.
\]
From (\ref{vVret2}) it follows that if the two systems begin at the same
corresponding points in configuration space -- that is, if $(q_{1}^{\prime
}(t_{i}),q_{2}^{\prime}(t_{i}))=(Q_{1}^{\prime}(t_{i}),Q_{2}^{\prime}(t_{i}))$
-- then the respective displacements over the time periods $(t_{i},t)$ and
$(t_{i},t_{\mathrm{ret}}(t))$ will be equal:%
\begin{equation}
\delta q_{r}^{\prime}(t,t_{i})=\delta Q_{r}^{\prime}(t_{\mathrm{ret}}%
,t_{i})\ . \label{equiv}%
\end{equation}

The result (\ref{equiv}) shows the complete dynamical equivalence of the two
systems -- the field oscillator on expanding space with an effective
time-dependent mass $m=a^{3}$ and frequency $\omega=k/a$, and the standard
oscillator with constant initial mass $m_{i}=a_{i}^{3}$ and frequency
$\omega_{i}=k/a_{i}$ -- with the time $t$ for the first system replaced by the
retarded time $t_{\mathrm{ret}}(t)$ for the second system (provided the
respective coordinates $q_{r}$, $Q_{r}$ are rescaled to $q_{r}^{\prime}$,
$Q_{r}^{\prime}$).

There is also a one-to-one correspondence between the equilibrium states for
the two systems. At $t=t_{i}$ the wave functions coincide and one system will
be in equilibrium if and only if the other is. From (\ref{equiv}) it follows
that the field system on expanding space will be in equilibrium at time $t$ if
and only if the equivalent oscillator is in equilibrium at time
$t_{\mathrm{ret}}(t)$.

Let us show this explicitly. From the expressions (\ref{exact psi}) and
(\ref{exact psi SHO}) for the respective wave functions $\psi(q_{1},q_{2},t)$
and $\psi_{\mathrm{SHO}}(Q_{1},Q_{2},t)$ we find the relation%
\begin{equation}
\frac{g_{-}(t)}{\omega_{i}}\left(  \left\vert \psi(q_{1},q_{2},t)\right\vert
^{2}\right)  |_{q_{1}^{\prime}=a,\ q_{2}^{\prime}=b}=\frac{1}{m_{i}\omega_{i}%
}\left(  \left\vert \psi_{\mathrm{SHO}}(Q_{1},Q_{2},t_{\mathrm{ret}%
}(t))\right\vert ^{2}\right)  |_{Q_{1}^{\prime}=a,\ Q_{2}^{\prime}=b}\ ,
\end{equation}
with each side evaluated at corresponding rescaled points $(a,b)$. The
left-hand side is the equilibrium distribution $\rho_{\mathrm{QT}}^{\prime
}(q_{1}^{\prime},q_{2}^{\prime},t)$ for the rescaled field variables, while
the right-hand side is the equilibrium distribution $\rho_{\mathrm{SHO\ QT}%
}^{\prime}(Q_{1}^{\prime},Q_{2}^{\prime},t_{\mathrm{ret}}(t))$ for the
rescaled oscillator. Thus we have%
\begin{equation}
\rho_{\mathrm{QT}}^{\prime}(q_{1}^{\prime},q_{2}^{\prime},t)=\rho
_{\mathrm{SHO\ QT}}^{\prime}(Q_{1}^{\prime},Q_{2}^{\prime},t_{\mathrm{ret}%
}(t)) \label{A}%
\end{equation}
(where it is understood that the two sides are evaluated at corresponding
points). If we assume that the initial -- generally nonequilibrium --
distributions for the two systems are equal, $\rho(q_{1},q_{2},t_{i}%
)=\rho_{\mathrm{SHO}}(Q_{1},Q_{2},t_{i})$, then since the initial rescalings
coincide the initial rescaled distributions will also be equal: $\rho^{\prime
}(q_{1}^{\prime},q_{2}^{\prime},t_{i})=\rho_{\mathrm{SHO}}^{\prime}%
(Q_{1}^{\prime},Q_{2}^{\prime},t_{i})$. From the correspondence (\ref{equiv})
between the rescaled trajectories it then follows that%
\begin{equation}
\rho^{\prime}(q_{1}^{\prime},q_{2}^{\prime},t)=\rho_{\mathrm{SHO}}^{\prime
}(Q_{1}^{\prime},Q_{2}^{\prime},t_{\mathrm{ret}}(t)) \label{B}%
\end{equation}
at all times $t$ -- that is, the rescaled density for the field system at time
$t$ is equal to the rescaled density for the oscillator at the retarded time
$t_{\mathrm{ret}}(t)$. From (\ref{A}) and (\ref{B}) we may write%
\[
\frac{\rho^{\prime}(q_{1}^{\prime},q_{2}^{\prime},t)}{\rho_{\mathrm{QT}%
}^{\prime}(q_{1}^{\prime},q_{2}^{\prime},t)}=\frac{\rho_{\mathrm{SHO}}%
^{\prime}(Q_{1}^{\prime},Q_{2}^{\prime},t_{\mathrm{ret}}(t))}{\rho
_{\mathrm{SHO\ QT}}^{\prime}(Q_{1}^{\prime},Q_{2}^{\prime},t_{\mathrm{ret}%
}(t))}\ .
\]
The field system will be in equilibrium at time $t$ (left-hand ratio equal to
one) if and only if the equivalent oscillator is in equilibrium at time
$t_{\mathrm{ret}}(t)$ (right-hand ratio equal to one).

The retarded time $t_{\mathrm{ret}}=t_{\mathrm{ret}}(t)$ is determined by
(\ref{tr}) for given functions $a(t)$, $g_{-}(t)$ on the interval $(t_{i},t)$.
For a radiation-dominated expansion, $a\propto t^{1/2}$, we have an exact
solution (\ref{gminus}) for $g_{-}(t)$ and the quantity $\Theta(t)$ has
already been evaluated (equation (\ref{Thetaeval})) so that we know the
function $t_{\mathrm{ret}}(t)=t_{i}+\Theta(t)/\omega_{i}$. Note that the
functions $g_{-}(t)$ and $t_{\mathrm{ret}}(t)$ depend on the wave number $k$
of the mode but are independent of the quantum state of the mode. Because of
the dynamical equivalence to the standard oscillator with retarded time, the
essential physics of our system on expanding space is determined by properties
of the function $t_{\mathrm{ret}}(t)$, which is in turn determined by the
function $g_{-}(t)$.

In the very short-time limit, $t=t_{i}+\Delta t$ with $\Delta t/t_{i}<<1$, we
have $m\simeq m_{i}$ and $g_{-}\simeq g_{-}(t_{i})=1/m_{i}$ and so we have
simply%
\[
t_{\mathrm{ret}}(t)\simeq t_{i}+\int_{t_{i}}^{t}dt^{\prime}=t\ .
\]
At very short times the retarded time $t_{\mathrm{ret}}$ reduces to real time
$t$.

\section{Freezing of quantum nonequilibrium in the far super-Hubble regime}

Two regimes are of particular significance: the far sub-Hubble and the far
super-Hubble limits. In Section 3 we introduced the parameter $\varepsilon
=\left(  t_{i}/a_{i}^{2}\right)  k^{2}$. Because%
\[
\varepsilon t=\frac{tt_{i}}{a_{i}^{2}}\left(  \frac{2\pi}{\lambda}\right)
^{2}=\frac{t^{2}}{a^{2}}\left(  \frac{2\pi}{\lambda}\right)  ^{2}=\left(
\frac{\pi H^{-1}}{a\lambda}\right)  ^{2}=\left(  \frac{\pi H^{-1}(t)}%
{\lambda_{\mathrm{phys}}(t)}\right)  ^{2}\ ,
\]
we may conveniently characterise the far sub-Hubble regime (with
$\lambda_{\mathrm{phys}}<<H^{-1}$) and the far super-Hubble regime (with
$\lambda_{\mathrm{phys}}>>H^{-1}$) by respective large or small values of
$\varepsilon t$.

In the far sub-Hubble regime we may therefore take $\varepsilon t_{i}>>1$ (in
which case we will also have $\varepsilon t>>1$ for all $t\geq t_{i}$). The
factor in square brackets in (\ref{gminus}) is then dominated by the term
$8\varepsilon t_{i}$ and so we have%
\begin{equation}
g_{-}(t)\simeq\frac{t_{i}}{a_{i}^{3}t} \label{g-subH}%
\end{equation}
in the far sub-Hubble limit. To recover the Minkowski limit we must consider
evolution over times $\Delta t<<H_{i}^{-1}=2t_{i}$ so that the scale factor
remains essentially constant. Thus, setting $m\simeq a_{i}^{3}$ and using
(\ref{g-subH}), at such a time $t_{i}+\Delta t$ the retarded time (\ref{tr})
will be%
\begin{equation}
t_{\mathrm{ret}}(t_{i}+\Delta t)\simeq t_{i}+\int_{t_{i}}^{t_{i}+\Delta
t}\frac{t^{\prime}}{t_{i}}dt^{\prime}=t_{i}+\Delta t+\frac{(\Delta t)^{2}%
}{2t_{i}}\simeq t_{i}+\Delta t\ .
\end{equation}
As expected, in the Minkowski limit the retarded time $t_{\mathrm{ret}}$
coincides with true time $t$. (This is a particular case of the short-time limit.)

For the far super-Hubble regime let us instead consider a time interval
$(t_{i},t_{f})$ during which $\varepsilon t<<1$. (We could simply set
$\varepsilon t_{f}<<1$, in which case we will also have $\varepsilon t<<1$ for
all $t\leq t_{f}$.) From (\ref{gminus}) we find that for $\varepsilon t<<1$
the function $g_{-}(t)$ takes the constant form%
\begin{equation}
g_{-}(t)\approx\frac{1}{a_{i}^{3}}=\frac{1}{m_{i}}\ . \label{g-super-H}%
\end{equation}
Inserting (\ref{g-super-H}) into (\ref{tr}), and writing $a=a_{i}%
(t/t_{i})^{1/2}$, we find that in the far super-Hubble regime the retarded
time is given by%
\begin{equation}
t_{\mathrm{ret}}(t)\simeq t_{i}+\int_{t_{i}}^{t}(t_{i}/t^{\prime}%
)^{3/2}dt^{\prime}=t_{i}+2t_{i}\left(  1-\sqrt{\frac{t_{i}}{t}}\right)  \ .
\end{equation}
In the short-time limit, $t=t_{i}+\Delta t$ with $\Delta t<<2t_{i}$, this
again reduces to%
\begin{equation}
t_{\mathrm{ret}}(t)\simeq t_{i}+\Delta t\ .
\end{equation}
However, in the long-time limit with $t_{f}>>t_{i}$ we now have%
\begin{equation}
t_{\mathrm{ret}}(t_{f})\simeq3t_{i}\ .
\end{equation}

This remarkable result may be stated as follows: in the far super-Hubble
regime, the long-time evolution of a field mode on an interval $(t_{i},t_{f})$
with $t_{f}>>t_{i}$ is equivalent to the time evolution of a standard harmonic
oscillator on the limited time interval $(t_{i},t_{\mathrm{ret}}%
(t_{f}))=(t_{i},3t_{i})$ (with appropriate rescaling of the coordinates). In
effect, \textit{the `equivalent standard oscillator' evolves over just one
Hubble time }$H^{-1}(t_{i})=2t_{i}$.

It is now very simple to deduce that, if the equivalent standard oscillator
has a relaxation timescale $\tau$ that is larger than $2t_{i}$ -- so that
equilibrium is not reached on the limited time interval $(t_{i},3t_{i})$ --
then the real field system will never reach equilibrium, not even for
$t_{f}>>t_{i}$ (for as long as the mode remains in the far super-Hubble
regime). Thus, in appropriate conditions, quantum nonequilibrium will be
`frozen' for super-Hubble modes.

We have reduced the question of relaxation on expanding space to the much
simpler question of relaxation for an equivalent standard oscillator. For the
standard (two-dimensional) oscillator it is straightforward to study the
relaxation timescale $\tau$ numerically. Analogous studies have already been
carried out for a particle in a two-dimensional box, for initial wave
functions that are superpositions of the first $M$ energy eigenstates
\cite{TRV12}. There it was found that the coarse-grained $H$-function $\bar
{H}$ decays approximately exponentially, $\bar{H}(t)\approx\bar{H}(t_{i}%
)\exp(-(t-t_{i})/\tau)$, with a timescale $\tau\propto1/M$ that scales
(approximately) inversely with $M$. We expect to find comparable behaviour for
the oscillator -- though with a somewhat different scaling of $\tau$ with $M$
for this different system. As the number $M$ of energy states in the
superposition increases, the relaxation timescale $\tau$ for the oscillator
will certainly decrease (owing to the increasing complexity of the de Broglie
velocity field). For $M$ larger than some critical value $M_{\max}$ we will
have $\tau\lesssim2t_{i}$ and we may deduce that the equivalent field system
will relax. If instead $M<M_{\max}$ we will have $\tau\gtrsim2t_{i}$ and the
field system will never reach equilibrium. (A detailed numerical study of
relaxation for the standard oscillator, and of the scaling of $\tau$ with $M$,
will be presented elsewhere \cite{AbV13}.)

The above conclusions agree at least qualitatively with the analysis given in
refs. \cite{AV08a,AVbook}. There it is shown that there is an upper bound on
the ratio $\left\langle \left\vert \delta q_{r}(t_{f})\right\vert
\right\rangle _{\mathrm{QT}}/\Delta_{r}(t_{f})$,%
\begin{equation}
\frac{\left\langle \left\vert \delta q_{r}(t_{f})\right\vert \right\rangle
_{\mathrm{QT}}}{\Delta_{r}(t_{f})}<4\sqrt{a_{f}^{3}\left\langle \hat{H}%
_{r}\right\rangle _{f}}\int_{t_{i}}^{t_{f}}dt\ \sqrt{\left\langle \hat{H}%
_{r}\right\rangle /a^{3}}\ , \label{UB}%
\end{equation}
where $\left\langle \left\vert \delta q_{r}(t_{f})\right\vert \right\rangle
_{\mathrm{QT}}$ is the (equilibrium) mean displacement of the trajectories
over the time interval $(t_{i},t_{f})$ and $\Delta_{r}(t_{f})\equiv
(1/2)(1/\Delta\pi_{r})$ is the characteristic lengthscale of the equilibrium
distribution in configuration space at time $t_{f}$ (where $\Delta\pi_{r}$ is
the quantum-theoretical spread for the canonical momentum operator $\hat{\pi
}_{r}$). In general, relaxation can occur only if the trajectories move over
distances that are at least comparable to $\Delta_{r}$. For super-Hubble modes
the right-hand side of (\ref{UB}) can be smaller than one -- in which case
relaxation will be suppressed, since most of the trajectories will not move
far enough for relaxation to occur \cite{AV08a,AVbook}. On the other hand,
clearly, the right-hand side of (\ref{UB}) can be large for a quantum state
with a sufficiently large mean Hamiltonian $\left\langle \hat{H}%
_{r}\right\rangle $, in which case no relaxation suppression can be deduced.

\section{Suppression of quantum noise at super-Hubble wavelengths}

The above results provide a mechanism whereby quantum noise can be suppressed
at super-Hubble wavelengths. If we assume that the initial nonequilibrium
distribution has a subquantum width, then under standard relaxation the
distribution evolves towards the Born rule and the width approaches the
standard quantum width. But in an expanding universe such relaxation can be
delayed -- in accordance with the retarded time $t_{\mathrm{ret}}(t)$ -- or
even completely frozen (in the far super-Hubble regime, as we saw in Section
5). In effect, as far as relaxation is concerned, over a time $t$ it is as if
only a time $t_{\mathrm{ret}}(t)<t$ has actually passed. Therefore in general
we expect that the actual width of the relaxing distribution will grow more
slowly and take longer to reach the quantum value -- or never reach it at all.

It is instructive to consider a numerical simulation that illustrates the
retardation effect.

We take an initial wave function that is a superposition%
\[
\psi(q_{1},q_{2},t_{i})=\frac{1}{\sqrt{M}}\sum_{n_{1}=0}^{\sqrt{M}-1}%
\sum_{n_{2}=0}^{\sqrt{M}-1}e^{i\theta_{n_{1}n_{2}}}\Phi_{n_{1}}(q_{1}%
)\Phi_{n_{2}}(q_{2})
\]
of instantaneous energy eigenstates $\Phi_{n_{1}}\Phi_{n_{2}}$ of the initial
Hamiltonian, with coefficients $c_{n_{1}n_{2}}(t_{i})=(1/\sqrt{M}%
)e^{i\theta_{n_{1}n_{2}}}$ of equal amplitude and with randomly-chosen initial
phases $\theta_{n_{1}n_{2}}$. (For simplicity the quantum numbers $n_{1}$,
$n_{2}$ are taken to have the same range; the number $M$ of modes is then
restricted to be the square of an integer.) As we saw in Section 3, the wave
function at time $t$ is then%
\[
\psi(q_{1},q_{2},t)=\frac{1}{\sqrt{M}}\sum_{n_{1}=0}^{\sqrt{M}-1}\sum
_{n_{2}=0}^{\sqrt{M}-1}e^{i\theta_{n_{1}n_{2}}}\psi_{n_{1}}(q_{1}%
,t)\psi_{n_{2}}(q_{2},t)\ ,
\]
where the exact solution for $\psi_{n}(q,t)$ is given by (\ref{psint}).

The quantum equilibrium distribution at time $t$ is given by $\rho
_{\mathrm{QT}}(q_{1},q_{2},t)=|\psi(q_{1},q_{2},t)|^{2}$. The actual
probability density at the initial time $t_{i}$ is taken to be%
\begin{equation}
\rho(q_{1},q_{2},t_{i})=|\Phi_{0}(q_{1})\Phi_{0}(q_{2})|^{2}=\frac{\omega
_{i}m_{i}}{\pi}e^{-m_{i}\omega_{i}q_{1}^{2}}e^{-m_{i}\omega_{i}q_{2}^{2}}~.
\end{equation}
This is equal to the equilibrium density for the quantum-theoretical ground
state $\Phi_{0}(q_{1})\Phi_{0}(q_{2})$. We choose this particular initial
distribution purely on grounds of simplicity. Clearly $\rho(q_{1},q_{2}%
,t_{i})\neq|\psi(q_{1},q_{2},t_{i})|^{2}$ and the initial state is far from
equilibrium. By calculating the de Broglie-Bohm trajectories $(q_{1}%
(t),q_{2}(t))$ numerically -- using de Broglie's equation of motion
(\ref{deB2D}) -- we may calculate the time evolution $\rho(q_{1},q_{2},t)$ of
the actual distribution and study whether or not it approaches the equilibrium
distribution $|\psi(q_{1},q_{2},t)|^{2}$ (on a coarse-grained level).

Because of the decreasing width of the solution (\ref{psint}), the support of
$|\psi(q_{1},q_{2},t)|^{2}$ in the $q_{1}-q_{2}$ plane shrinks with time. When
plotting the distributions it is therefore convenient to use the rescaled
variables $q_{r}^{\prime}=\sqrt{\omega_{i}/g_{-}(t)}q_{r}$ (with $g_{-}%
(t_{i})=1/m_{i}$ and where $g_{-}$ decreases with time). The equilibrium
probability density in the $q_{1}^{\prime}-q_{2}^{\prime}$ plane is then given
by%
\begin{equation}
\rho_{\mathrm{QT}}^{\prime}(q_{1}^{\prime},q_{2}^{\prime},t)=\frac{g_{-}%
(t)}{\omega_{i}}|\psi(q_{1},q_{2},t)|^{2}\ .
\end{equation}
In terms of the rescaled variables $q_{r}^{\prime}=\sqrt{m_{i}\omega_{i}}%
q_{r}$ at $t=t_{i}$ we have an initial nonequilibrium density%
\begin{equation}
\rho^{\prime}(q_{1}^{\prime},q_{2}^{\prime},t_{i})=\frac{1}{\omega_{i}m_{i}%
}\rho(q_{1},q_{2},t_{i})=\frac{1}{\pi}e^{-(q_{1}^{\prime})^{2}}e^{-(q_{2}%
^{\prime})^{2}}~.
\end{equation}
At later times $t$ the actual density in the $q_{1}^{\prime}-q_{2}^{\prime}$
plane is%
\begin{equation}
\rho^{\prime}(q_{1}^{\prime},q_{2}^{\prime},t)=\frac{g_{-}(t)}{\omega_{i}}%
\rho(q_{1},q_{2},t)\ .
\end{equation}

During a radiation-dominated expansion, a mode that begins with a super-Hubble
wavelength (that is, with a physical wavelength $\lambda_{\mathrm{phys}}%
(t_{i})>H_{i}^{-1}$) will enter the Hubble radius at a later time
$t_{\mathrm{enter}}$ that is determined by $\lambda_{\mathrm{phys}%
}(t_{\mathrm{enter}})=H^{-1}(t_{\mathrm{enter}})$. Thereafter the mode will
acquire a sub-Hubble wavelength (that is, $\lambda_{\mathrm{phys}}%
(t)<H^{-1}(t)$). Here $\lambda_{\mathrm{phys}}(t)=a(t)\lambda$, where the
comoving wavelength $\lambda$ is equal to the physical wavelength at a time
$t_{0}$ such that $a_{0}=1$ (often taken to be the time today).

We are particularly interested in modes that begin outside the Hubble radius.
Let us consider the evolution of such a mode during the entire super-Hubble
era -- that is, from an initial time $t_{i}$ until the time $t_{\mathrm{enter}%
}$. For the purposes of a numerical computation we have found it convenient to
take $t_{i}=10^{-4}$ and $t_{0}=1$. We then have $a_{i}=10^{-2}$. If we choose
$\lambda=0.2$ (or $k=2\pi/\lambda=10\pi$) then $\lambda_{\mathrm{phys}}%
(t_{i})=a_{i}\lambda=2\times10^{-3}$ and $H_{i}^{-1}=2t_{i}=2\times10^{-4}$.
At the initial time the mode is outside the Hubble radius by one order of
magnitude. Mode entry occurs at $t_{\mathrm{enter}}=10^{-2}$. (These chosen
values are not intended to have any particular cosmological significance, they
are for numerical convenience and illustration only.)

The equivalence theorem of Section 4 tells us that the time evolution of the
real system on expanding space, over the time interval $(t_{i}%
,\ t_{\mathrm{enter}})$, may be obtained by evolving the equivalent standard
oscillator (with the same initial conditions for the wave function and
nonequilibrium distribution) over the time interval $(t_{i},\ t_{\mathrm{ret}%
}(t_{\mathrm{enter}}))$ -- where $t_{\mathrm{ret}}(t)$ is the retarded time
corresponding to real time $t$. The required values of $t_{\mathrm{ret}}$ may
be obtained from the analytical result (\ref{Thetaeval}) for $\Theta$, where
$t_{\mathrm{ret}}(t)=t_{i}+\Theta(t)/\omega_{i}$. A plot of the required
function $t_{\mathrm{ret}}(t)$ -- for the above values of the parameters
$t_{i}$, $a_{i}$, $k$ and over the time interval $(t_{i},\ t_{\mathrm{enter}%
})$ -- is given in Figure 1. (For this plot the parameter $\varepsilon$ is
equal to $100\pi^{2}$.)%

%TCIMACRO{\FRAME{ftbpFU}{4.0007in}{3.3993in}{0pt}{\Qcb{Plot of the retarded
%time $t_{\QTR{rm}{ret}}=t_{\QTR{rm}{ret}}(t)$ (solid line) for $t$ on the
%interval $(t_{i},\ t_{\QTR{rm}{enter}})$. The dotted line is a plot of real
%time $t$. The function $t_{\QTR{rm}{ret}}(t)$ is given in terms of $\Theta(t)$
%by $t_{\QTR{rm}{ret}}(t)=t_{i}+\Theta(t)/\omega_{i}$ where $\Theta(t)$ is
%given by equation (\ref{Thetaeval}). We have chosen parameters $t_{i}=10^{-4}%
%$, $a_{i}=10^{-2}$ and $k=10\pi$ (so that $\varepsilon=100\pi^{2}$).}}%
%{}{figi.jpg}{\special{ language "Scientific Word";  type "GRAPHIC";
%width 4.0007in;  height 3.3993in;  depth 0pt;  original-width 9.934in;
%original-height 8.4267in;  cropleft "0";  croptop "1";  cropright "1";
%cropbottom "0";  filename 'FigI.jpg';file-properties "XNPEU";}} }%
%BeginExpansion
\begin{figure}
\begin{center}
\includegraphics[width=0.7\textwidth]{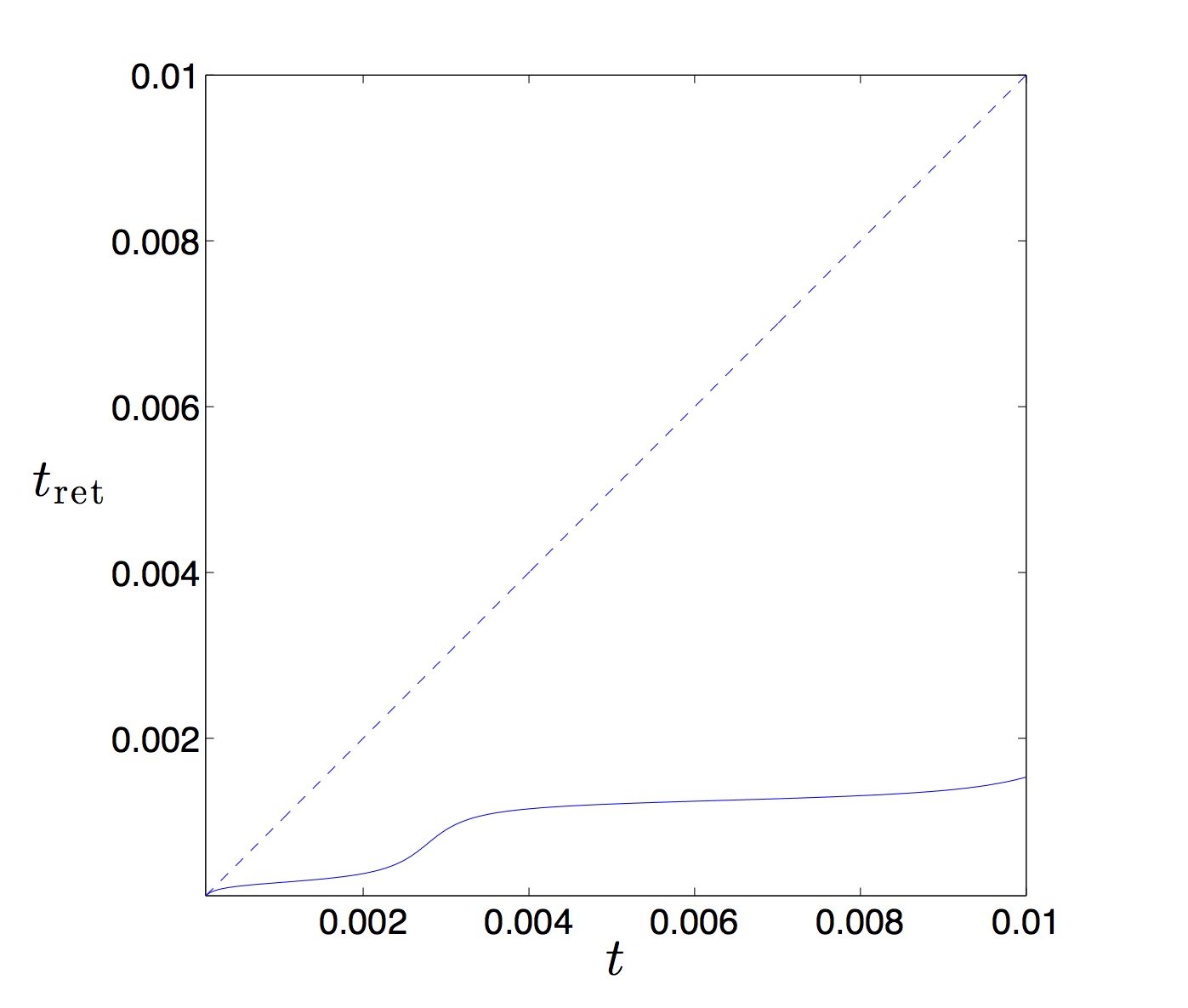}
\caption{Plot of the retarded time $t_{\mathrm{ret}}=t_{\mathrm{ret}}(t)$
(solid line) for $t$ on the interval $(t_{i},\ t_{\mathrm{enter}})$. The
dotted line is a plot of real time $t$. The function $t_{\mathrm{ret}}(t)$ is
given in terms of $\Theta(t)$ by $t_{\mathrm{ret}}(t)=t_{i}+\Theta
(t)/\omega_{i}$ where $\Theta(t)$ is given by equation (\ref{Thetaeval}). We
have chosen parameters $t_{i}=10^{-4}$, $a_{i}=10^{-2}$ and $k=10\pi$ (so that
$\varepsilon=100\pi^{2}$).}%
\end{center}
\end{figure}
%EndExpansion

The time evolution of the equivalent standard oscillator may be obtained by
straightforward numerical simulation. We employ the `backtracking' method of
ref. \cite{VW05}, which uses the conserved ratio $\rho^{\prime}/\rho
_{\mathrm{QT}}^{\prime}$ along trajectories to construct $\rho^{\prime}$ on a
uniform grid at each time $t$. Our grid consists of $1000\times1000$ points.
We impose a precision of $0.01$ on the backtracked trajectories (compared to a
linear scale of $\sim10$ for the support of the distributions, and where
$\rho_{\mathrm{QT}}^{\prime}$ displays structure on lengthscales down to
$\sim1$). The evolving density $\rho^{\prime}$ develops an extremely irregular
fine-grained structure, with rapid variations over very short distances (cf.
figure 6 of ref. \cite{VW05}). The density may be averaged over
coarse-graining cells, with a coarse-grained value assigned to the centre of
each cell. It is convenient to plot a `smoothed' density $\tilde{\rho}%
^{\prime}$ obtained by coarse-graining with overlapping cells \cite{VW05}%
.\footnote{The plots in Figures 2 and 3 employ $96\times96$ overlapping cells
each with $50\times50$ grid points. The cells have side $\varepsilon=0.5$. For
a given cell, shifting it along either axis by a distance equal to $20\%$ of
$\varepsilon$ generates a neighbouring cell.}

In Figure 2 we show the result of such a simulation for the case of 25 modes
($M=25$). The top row shows the initial (smoothed) actual distribution
$\tilde{\rho}^{\prime}(t_{i})$ on the left-hand side and the initial
(smoothed) equilibrium distribution $\tilde{\rho}_{\mathrm{QT}}^{\prime}%
(t_{i})$ on the right-hand side. The support of $\tilde{\rho}^{\prime}(t_{i})$
is considerably narrower than the support of $\tilde{\rho}_{\mathrm{QT}%
}^{\prime}(t_{i})$. The second row shows the (smoothed) distributions at an
intermediate retarded time $t_{\mathrm{ret}}=t_{\mathrm{ret}}%
(0.5t_{\mathrm{enter}})=1.21\times10^{-3}$, while the third row shows these at
$t_{\mathrm{ret}}=t_{\mathrm{ret}}(t_{\mathrm{enter}})=1.53\times10^{-3}$. The
three times $t_{i}$, $t_{\mathrm{ret}}(0.5t_{\mathrm{enter}})$,
$t_{\mathrm{ret}}(t_{\mathrm{enter}})$ for the equivalent oscillator
correspond to the times $t_{i}$, $0.5t_{\mathrm{enter}}$, $t_{\mathrm{enter}}$
for the real system on expanding space. As is plain from the figure, the
support of $\tilde{\rho}^{\prime}$ spreads out -- at least initially -- while
the support of $\tilde{\rho}_{\mathrm{QT}}^{\prime}$ remains about the same
(with the rescaled coordinates). However, over the time interval considered,
the support of $\tilde{\rho}^{\prime}$ remains significantly narrower than the
support of $\tilde{\rho}_{\mathrm{QT}}^{\prime}$. There has clearly been only
a partial relaxation towards equilibrium (as will be quantified below using
the coarse-grained $H$-function).%

%TCIMACRO{\FRAME{ftbpFU}{3.4778in}{4.4491in}{0pt}{\Qcb{Time evolution of
%nonequilibrium on expanding space, for a superposition of 25 modes. Results
%for the interval $(t_{i},t_{\QTR{rm}{enter}})$ are obtained by evolving the
%equivalent oscillator over the retarded interval $(t_{i}%
%,t_{\QTR{rm}{ret}}(t_{\QTR{rm}{enter}}))$. The (smoothed) actual distribution
%$\tilde{\rho}^{\prime}$ is displayed in the left column, the (smoothed)
%equilibrium distribution $\tilde{\rho}_{\QTR{rm}{QT}}^{\prime}$ in the right
%column. The top row shows the distributions at the initial time $t_{i}$, the
%second row at an intermediate retarded time
%$t_{\QTR{rm}{ret}}(0.5t_{\QTR{rm}{enter}})$, and the third row at
%$t_{\QTR{rm}{ret}}(t_{\QTR{rm}{enter}})$. The support of $\tilde{\rho}%
%^{\prime}$ remains significantly narrower than the support of $\tilde{\rho
%}_{\QTR{rm}{QT}}^{\prime}$.}}{}{figii.jpg}%
%width 3.4778in;  height 4.4491in;  depth 0pt;  original-width 6.8272in;
%original-height 10.9155in;  cropleft "0";  croptop "1";  cropright "1";
%cropbottom "0";  filename 'FigII.jpg';file-properties "XNPEU";}} }%
%BeginExpansion
\begin{figure}
\begin{center}
\includegraphics[width=\textwidth]{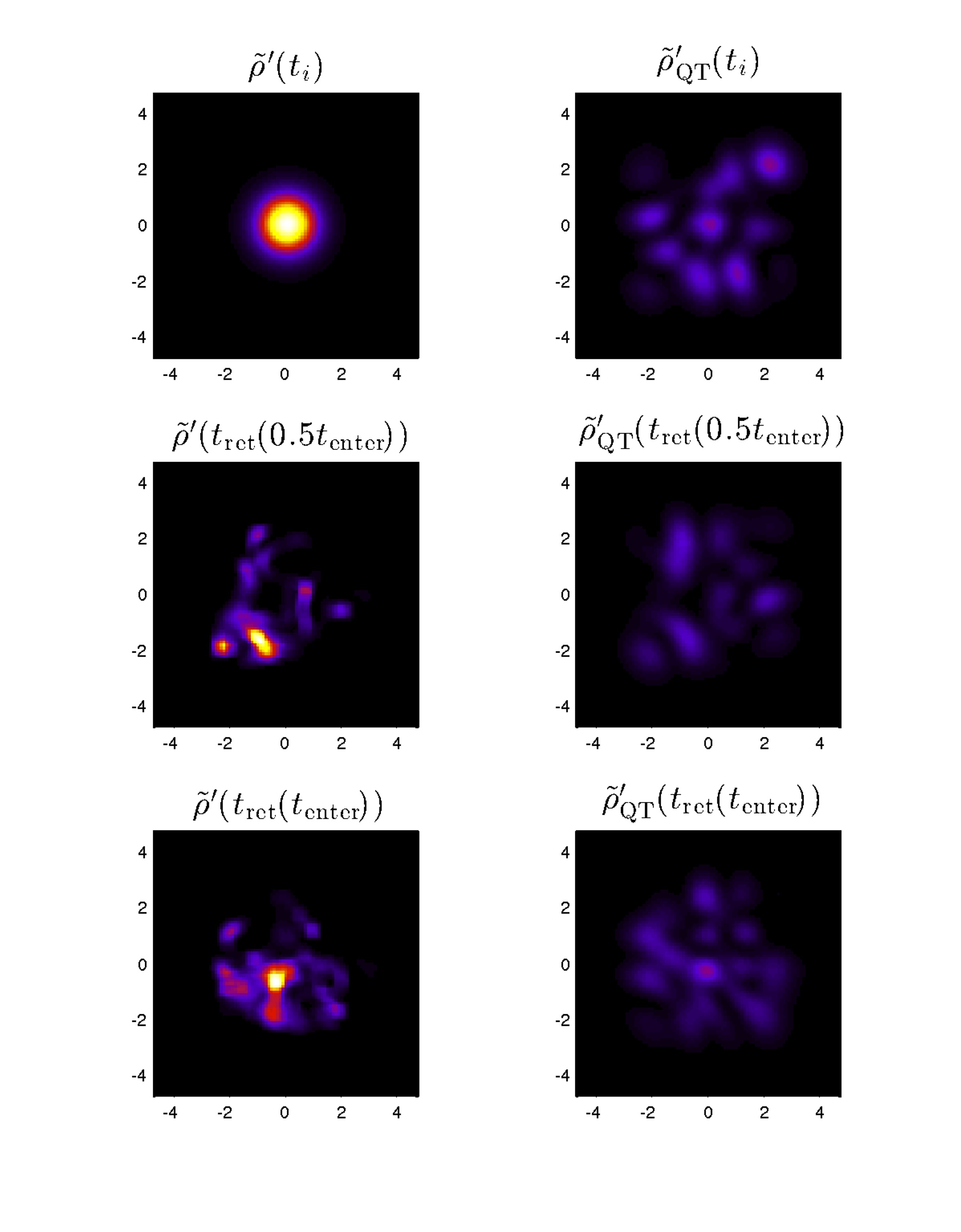}
\caption{Time evolution of nonequilibrium on expanding space, for a
superposition of 25 modes. Results for the interval $(t_{i},t_{\mathrm{enter}%
})$ are obtained by evolving the equivalent oscillator over the retarded
interval $(t_{i},t_{\mathrm{ret}}(t_{\mathrm{enter}}))$. The (smoothed) actual
distribution $\tilde{\rho}^{\prime}$ is displayed in the left column, the
(smoothed) equilibrium distribution $\tilde{\rho}_{\mathrm{QT}}^{\prime}$ in
the right column. The top row shows the distributions at the initial time
$t_{i}$, the second row at an intermediate retarded time $t_{\mathrm{ret}%
}(0.5t_{\mathrm{enter}})$, and the third row at $t_{\mathrm{ret}%
}(t_{\mathrm{enter}})$. The support of $\tilde{\rho}^{\prime}$ remains
significantly narrower than the support of $\tilde{\rho}_{\mathrm{QT}}%
^{\prime}$.}%
\end{center}
\end{figure}
%EndExpansion

Let us contrast this result with a simulation for the same standard
oscillator, with the same initial conditions, but evolved up to a time
$t=t_{\mathrm{enter}}=10^{-2}$. Physically, this would correspond to the time
evolution of the real system with no spatial expansion (that is, with $a=1$
for all $t$ so that $t_{\mathrm{ret}}(t)=t$). The results are shown in Figure
3. The first row shows the same initial conditions as before. The second and
third rows show the (smoothed) distributions at the respective times
$t=0.5t_{\mathrm{enter}}$ and $t=t_{\mathrm{enter}}$. The results speak for
themselves. Already by $t=0.5t_{\mathrm{enter}}$ the actual distribution
$\tilde{\rho}^{\prime}$ has a support that is only slightly narrower than the
support of $\tilde{\rho}_{\mathrm{QT}}^{\prime}$. At $t=t_{\mathrm{enter}}$
there is little discernible difference between the distributions $\tilde{\rho
}^{\prime}$ and $\tilde{\rho}_{\mathrm{QT}}^{\prime}$ -- not only in terms of
the extent of their support but also as regards detailed features. There has
clearly been an almost complete relaxation to equilibrium.%

%TCIMACRO{\FRAME{ftbpFU}{3.4769in}{4.4491in}{0pt}{\Qcb{Time evolution of the
%same initial state as in Figure 2 but with no spatial expansion. The results
%for $(t_{i},t_{\QTR{rm}{enter}})$ are now obtained simply by evolving the
%standard oscillator over $(t_{i},t_{\QTR{rm}{enter}})$. The top row again
%shows the (smoothed) distributions at the initial time $t_{i}$, the second row
%at the intermediate time $0.5t_{\QTR{rm}{enter}}$, and the third row at
%$t_{\QTR{rm}{enter}}$. Already at $t=0.5t_{\QTR{rm}{enter}}$ the actual
%distribution $\tilde{\rho}^{\prime}$ has a support that is only slightly
%narrower than the support of $\tilde{\rho}_{\QTR{rm}{QT}}^{\prime}$. At
%$t=t_{\QTR{rm}{enter}}$ there is little discernible difference between
%$\tilde{\rho}^{\prime}$ and $\tilde{\rho}_{\QTR{rm}{QT}}^{\prime}$ --
%relaxation is almost complete.}}{}{figiii.jpg}%
%width 3.4769in;  height 4.4491in;  depth 0pt;  original-width 6.8272in;
%original-height 10.9155in;  cropleft "0";  croptop "1";  cropright "1";
%cropbottom "0";  filename 'FigIII.jpg';file-properties "XNPEU";}} }%
%BeginExpansion
\begin{figure}
\begin{center}
\includegraphics[width=\textwidth]%
{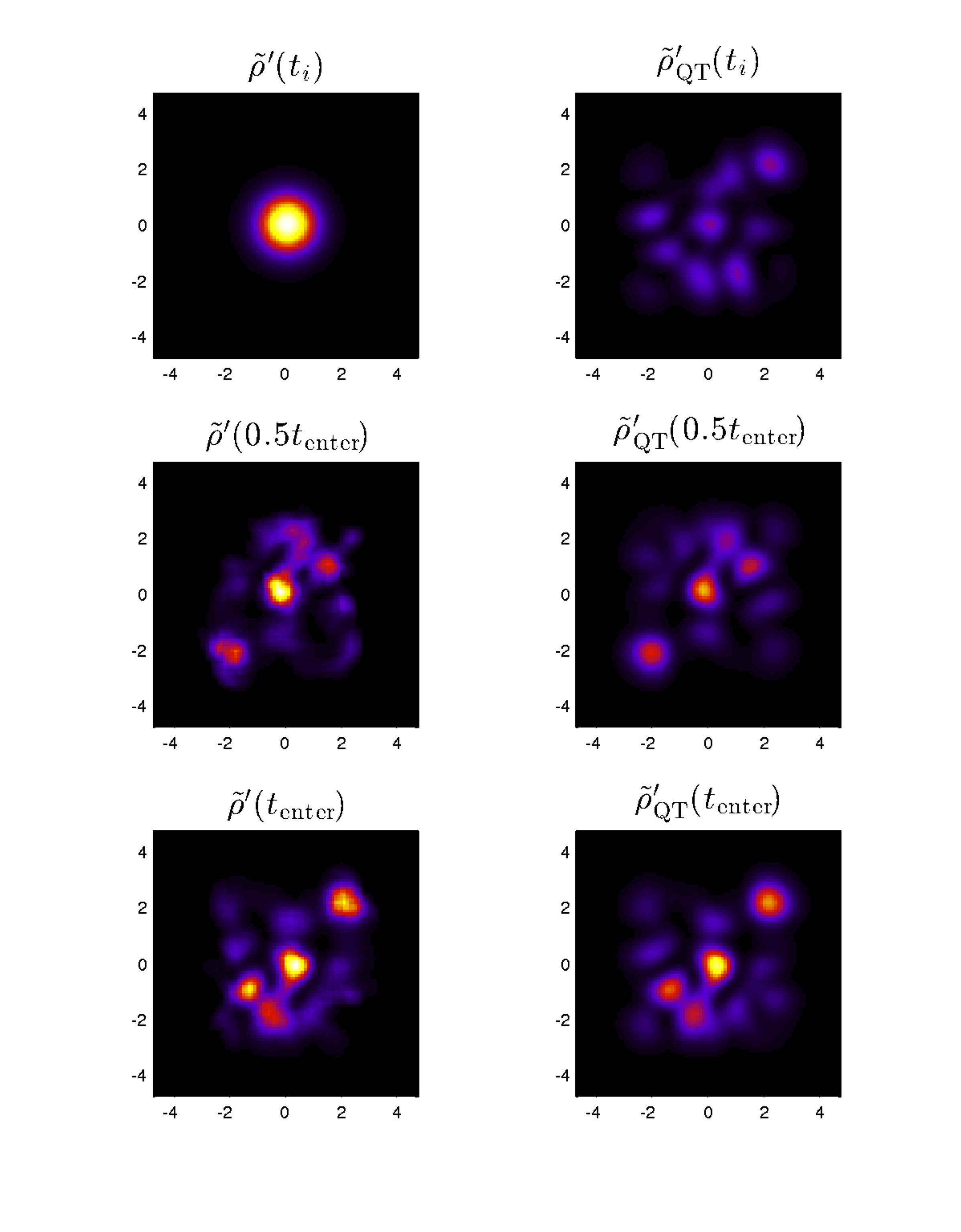}%
\caption{Time evolution of the same initial state as in Figure 2 but with no
spatial expansion. The results for $(t_{i},t_{\mathrm{enter}})$ are now
obtained simply by evolving the standard oscillator over $(t_{i}%
,t_{\mathrm{enter}})$. The top row again shows the (smoothed) distributions at
the initial time $t_{i}$, the second row at the intermediate time
$0.5t_{\mathrm{enter}}$, and the third row at $t_{\mathrm{enter}}$. Already at
$t=0.5t_{\mathrm{enter}}$ the actual distribution $\tilde{\rho}^{\prime}$ has
a support that is only slightly narrower than the support of $\tilde{\rho
}_{\mathrm{QT}}^{\prime}$. At $t=t_{\mathrm{enter}}$ there is little
discernible difference between $\tilde{\rho}^{\prime}$ and $\tilde{\rho
}_{\mathrm{QT}}^{\prime}$ -- relaxation is almost complete.}%
\end{center}
\end{figure}
%EndExpansion

The approach to equilibrium may be quantified using the coarse-grained
$H$-function%
\begin{equation}
\bar{H}=\int\int dq_{1}^{\prime}dq_{2}^{\prime}\ \bar{\rho}^{\prime}\ln
(\bar{\rho}^{\prime}/\bar{\rho}_{\mathrm{QT}}^{\prime})\ ,
\end{equation}
where $\bar{\rho}^{\prime}$, $\bar{\rho}_{\mathrm{QT}}^{\prime}$ are obtained
by averaging $\rho^{\prime}$, $\rho_{\mathrm{QT}}^{\prime}$ over
(non-overlapping) coarse-graining cells. As we recalled in Section 1, this
function obeys a coarse-graining $H$-theorem \cite{AV91a,AV92} and provides a
convenient measure of relaxation. For the above two simulations, a plot of
$\ln\bar{H}$ as a function of time $t$ is shown in Figure 4. In both cases
real time runs from $t=t_{i}$ up to $t=t_{\mathrm{enter}}$. In the case with
no spatial expansion the $\bar{H}$-curve has a larger (negative)\ slope and
ends with a smaller value -- the relaxation proceeds more quickly and the
final distribution comes considerably closer to equilibrium. (The early part
of the time evolution shows a clear exponential decay, which then appears to
tail off somewhat.\footnote{Here we employ $20\times20$ non-overlapping
coarse-graining cells each containing $50\times50$ grid points. The error bars
are obtained by running the same simulation with different grids so as to
obtain different samples of the highly fine-grained function $\rho^{\prime}$.})%

%TCIMACRO{\FRAME{ftbpFU}{4.2895in}{2.1327in}{0pt}{\Qcb{Plots of $\ln\bar{H}$
%against time $t$, with spatial expansion (upper curve)\ and with no spatial
%expansion (lower curve). Real time runs from $t=t_{i}$ up to
%$t=t_{\QTR{rm}{enter}}$. The lower curve has a larger (negative)\ slope and
%ends with a smaller value. With no spatial expansion there is faster
%relaxation and the final distribution comes considerably closer to
%equilibrium. The difference between the two $\bar{H}$-curves quantifies the
%suppression of relaxation on expanding space in the super-Hubble regime.}}%
%{}{figiv.jpg}{\special{ language "Scientific Word";  type "GRAPHIC";
%width 4.2895in;  height 2.1327in;  depth 0pt;  original-width 21.3118in;
%original-height 10.518in;  cropleft "0";  croptop "1";  cropright "1";
%cropbottom "0";  filename 'FigIV.jpg';file-properties "XNPEU";}} }%
%BeginExpansion
\begin{figure}
\begin{center}
\includegraphics[width=0.9\textwidth]%
{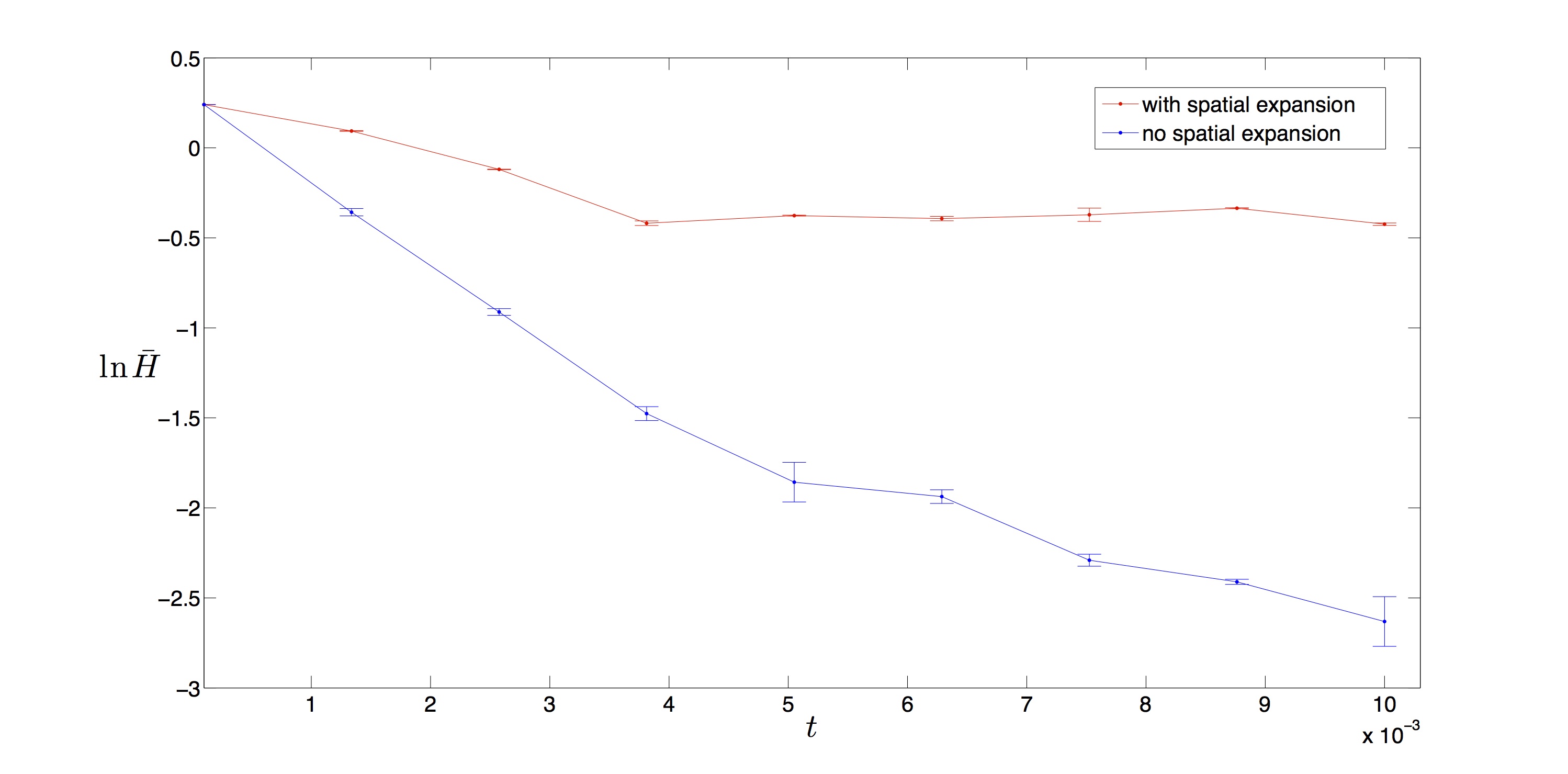}%
\caption{Plots of $\ln\bar{H}$ against time $t$, with spatial expansion (upper
curve)\ and with no spatial expansion (lower curve). Real time runs from
$t=t_{i}$ up to $t=t_{\mathrm{enter}}$. The lower curve has a larger
(negative)\ slope and ends with a smaller value. With no spatial expansion
there is faster relaxation and the final distribution comes considerably
closer to equilibrium. The difference between the two $\bar{H}$-curves
quantifies the suppression of relaxation on expanding space in the
super-Hubble regime.}%
\end{center}
\end{figure}
%EndExpansion

The contrast between Figures 2 and 3 -- quantified by the different $\bar{H}%
$-curves in Figure 4 -- provides a graphic illustration of our mechanism for
the suppression of quantum noise at super-Hubble wavelengths. The effect of
the spatial expansion is to \textit{retard relaxation in the super-Hubble
regime}. For an initial nonequilibrium distribution with a subquantum width,
at later times the width can remain subquantum -- even though, over the same
time interval, almost complete relaxation would have occurred if space had not
been expanding.

\section{Mechanism for a large-scale power deficit in the CMB}

We have demonstrated a mechanism for the suppression of quantum noise at
super-Hubble wavelengths in a radiation-dominated universe. It has been
suggested that such a mechanism would generate a large-scale power deficit in
the CMB in a cosmology with a radiation-dominated pre-inflationary phase
\cite{AV07,AV08a,AV10}. Relaxation suppression could have occurred in the
pre-inflationary era, resulting in a deficit in the inflationary spectrum
above some large comoving wavelength $\lambda_{\mathrm{c}}$. However, no
estimate was given for the value of $\lambda_{\mathrm{c}}$.

The existence of a large-scale power deficit in the CMB has recently been
confirmed by the \textit{Planck} satellite \cite{PlanckXV}. The reported
statistical significance is not high (in the range 2.5--3$\sigma$). It is
therefore quite possible that the primordial power spectrum for a theoretical
ensemble of skies is not itself anomalous, and that we have simply observed a
chance fluctuation for our single sky (see Section 7.1). Even so, it is worth
exploring models that predict a genuine deficit in the primordial spectrum, so
as to better assess the significance of what has been observed.

We now provide a simple estimate for $\lambda_{\mathrm{c}}$ which depends
essentially on the number $N$ of inflationary e-folds and on the inflationary
reheating temperature $T_{\mathrm{end}}$. The allowed values for $N$ and
$T_{\mathrm{end}}$ are consistent with a cutoff $\lambda_{\mathrm{c}}$
corresponding to the scale of the observed power deficit. On the other hand,
if $N$ is very large then our cutoff $\lambda_{\mathrm{c}}$ will be far too
big to yield an observable effect on the CMB.

There are of course other possible effects that could contribute to the
observed deficit (and perhaps account for it in full). For example, a deficit
could arise from a period of `fast rolling' for the inflaton field around the
beginning of the last 65 e-folds of inflation \cite{CL03}. A
radiation-dominated pre-inflationary phase can also yield corrections to the
quantum vacuum state during inflation, resulting in a loss of power at low $l$
\cite{PK07,WN08}. Our main concern here is to show that our mechanism for
quantum noise suppression on large scales could have implications for the CMB.
As will be discussed further in Section 8, the development of a detailed
cosmological model and comparisons with other possible effects are left for
future work.

\subsection{The CMB in the low-$l$ region\newline}

We first briefly review the standard treatment of the CMB at large angular scales.

The temperature anisotropy $\Delta T(\theta,\phi)\equiv T(\theta,\phi)-\bar
{T}$ of the CMB sky (where $\bar{T}$ is the average over the sky) may be
decomposed into spherical harmonics,%
\begin{equation}
\frac{\Delta T(\theta,\phi)}{\bar{T}}=\sum_{l=2}^{\infty}\sum_{m=-l}%
^{+l}a_{lm}Y_{lm}(\theta,\phi)\ . \label{har}%
\end{equation}
It is usual to regard $T(\theta,\phi)$ as a single realisation of a stochastic
process such that the marginal probability distribution for each coefficient
$a_{lm}$ is independent of $m$. This will be true if the probability
distribution for $T(\theta,\phi)$ -- over a theoretical `ensemble of skies' --
is rotationally invariant. The predicted angular power spectrum%
\begin{equation}
C_{l}\equiv\left\langle \left\vert a_{lm}\right\vert ^{2}\right\rangle
\end{equation}
then depends only on $l$ (where $\left\langle ...\right\rangle $ denotes an
average over the theoretical ensemble). The quantity%
\begin{equation}
C_{l}^{\mathrm{sky}}\equiv\frac{1}{2l+1}\sum_{m=-l}^{+l}\left\vert
a_{lm}\right\vert ^{2}%
\end{equation}
is constructed from measurements on a single sky and satisfies $\left\langle
C_{l}^{\mathrm{sky}}\right\rangle =C_{l}$. Thus $C_{l}^{\mathrm{sky}}$ gives
an unbiased estimate of $C_{l}$. It has a cosmic variance $\Delta
C_{l}^{\mathrm{sky}}/C_{l}=\sqrt{2/2l+1}$. (In practice, of course, the CMB
data contain additional noise and errors that must be accounted for.)

The temperature anisotropy is generated from primordial curvature
perturbations $\mathcal{R}_{\mathbf{k}}\equiv(1/4)\left(  a/k\right)
^{2}\,^{(3)}R_{\mathbf{k}}$ (where $^{(3)}R_{\mathbf{k}}$ is the Fourier
component of the spatial curvature scalar on comoving hypersurfaces) in
accordance with the formula \cite{LR99}%
\begin{equation}
a_{lm}=\frac{i^{l}}{2\pi^{2}}\int d^{3}\mathbf{k}\ \mathcal{T}(k,l)\mathcal{R}%
_{\mathbf{k}}Y_{lm}(\mathbf{\hat{k}})\ , \label{alm}%
\end{equation}
where the transfer function $\mathcal{T}(k,l)$ encodes the relevant
astrophysical processes.

If the probability distribution for $\mathcal{R}_{\mathbf{k}}$ is
translationally invariant it follows that $\left\langle \mathcal{R}%
_{\mathbf{k}}\mathcal{R}_{\mathbf{k%
%TCIMACRO{\U{b4}}%
%BeginExpansion
\acute{}%
%EndExpansion
}}^{\ast}\right\rangle =\delta_{\mathbf{kk}%
%TCIMACRO{\U{b4}}%
%BeginExpansion
\acute{}%
%EndExpansion
}\left\langle \left\vert \mathcal{R}_{\mathbf{k}}\right\vert ^{2}\right\rangle
$. From (\ref{alm}) one then obtains the expression%
\begin{equation}
C_{l}=\frac{1}{2\pi^{2}}\int_{0}^{\infty}\frac{dk}{k}\ \mathcal{T}%
^{2}(k,l)\mathcal{P}_{\mathcal{R}}(k) \label{Cl2}%
\end{equation}
for the angular power spectrum in terms of the primordial power spectrum%
\begin{equation}
\mathcal{P}_{\mathcal{R}}(k)\equiv\frac{4\pi k^{3}}{V}\left\langle \left\vert
\mathcal{R}_{\mathbf{k}}\right\vert ^{2}\right\rangle \label{PPS}%
\end{equation}
(with $V$ a normalisation volume). This provides a link between the statistics
of the primordial perturbations and the observed features in the CMB. The data
for $C_{l}$ are consistent with an approximately scale-free spectrum
$\mathcal{P}_{\mathcal{R}}(k)\approx\mathrm{const.}$.

At large angular scales -- that is, for small values of $l$ (say $l\lesssim
20$) -- the angular power spectrum $C_{l}$ is dominated by the Sachs-Wolfe
effect. In this region the square of the transfer function takes the simple
form \cite{LL00}%
\begin{equation}
\mathcal{T}^{2}(k,l)=\pi H_{0}^{4}j_{l}^{2}(2k/H_{0})\ ,
\end{equation}
where $H_{0}$ is the Hubble parameter today. From (\ref{Cl2}) we then have%
\begin{equation}
C_{l}=\frac{H_{0}^{4}}{2\pi}\int_{0}^{\infty}\frac{dk}{k}\ j_{l}^{2}%
(2k/H_{0})\mathcal{P}_{\mathcal{R}}(k)\ . \label{Clowl}%
\end{equation}
For an exactly scale-invariant spectrum, $\mathcal{P}_{\mathcal{R}%
}(k)=\mathrm{const}.$, this yields $C_{l}\propto1/l(l+1)$. (The integrated
Sachs-Wolfe effect will cause a small rise in the value of $l(l+1)C_{l}$ at
very small $l$.)

There were suggestions that the CMB data from the \textit{WMAP} satellite
contained anomalously low power at small $l$. Such claims were, however,
controversial. (For a review and critical assessment based on the seven-year
\textit{WMAP} data see ref. \cite{B11}.) Recently, the anomaly has been
confirmed to exist in data from the \textit{Planck} satellite \cite{PlanckXV}.

The \textit{Planck} team report a power deficit of 5--10\% in the region
$l\lesssim40$, with a statistical significance in the range 2.5--3$\sigma$
(depending on the estimator that is used). While the statistical significance
is not high, the \textit{Planck} team have noted the importance of finding a
theoretical model that predicts a low-$l$ deficit.

If the reported power deficit is not due to inadequate data processing or to
some local astrophysical effect then it must be primordial in origin. It might
be regarded as a mere random fluctuation for our single sky. Otherwise, it
reflects a genuine anomaly in the primordial power spectrum $\mathcal{P}%
_{\mathcal{R}}(k)$ for the theoretical ensemble. To explain such an anomaly
would presumably require a modification of the standard inflationary scenario
-- and perhaps some new physics.

\subsection{Inflation with early quantum nonequilibrium}

Inflationary cosmology predicts a curvature perturbation $\mathcal{R}%
_{\mathbf{k}}$ that may be obtained from the simple formula \cite{LL00}%
\begin{equation}
\mathcal{R}_{\mathbf{k}}=-\left[  \frac{H}{\dot{\phi}_{0}}\phi_{\mathbf{k}%
}\right]  _{t=t_{\ast}(k)}\ . \label{form}%
\end{equation}
Here $H$ is the (approximately constant) Hubble parameter of the inflating
universe, while $\phi_{0}$ and $\phi$ respectively denote the spatially
homogeneous and inhomogeneous parts of the inflaton field. The right-hand side
is evaluated at a time $t_{\ast}(k)$ taken to be a few $e$-folds after the
exponentially-expanding physical wavelength $\lambda_{\mathrm{phys}}%
=a(2\pi/k)$ of the mode exits the Hubble radius. The inflaton perturbation
$\phi$ is defined on a spatially flat slicing, while the curvature
perturbation $\mathcal{R}_{\mathbf{k}}$ is defined on the comoving slicing.
Thus (\ref{form}) relates quantities defined on different
slicings.\footnote{Note that (\ref{form}) becomes singular if one literally
takes the slow-roll limit $\dot{\phi}_{0}\rightarrow0$. The formula should be
understoood to be valid in the near-de Sitter regime and not for a strict de
Sitter expansion. This simple treatment suffices for our purposes.}

In an ideal Bunch-Davies vacuum the inflaton perturbations $\phi_{\mathbf{k}}$
will have (at time $t_{\ast}(k)$) a quantum-theoretical variance%
\begin{equation}
\left\langle |\phi_{\mathbf{k}}|^{2}\right\rangle _{\mathrm{QT}}=\frac
{V}{2(2\pi)^{3}}\frac{H^{2}}{k^{3}} \label{B-D}%
\end{equation}
and a scale-free power spectrum%
\begin{equation}
\mathcal{P}_{\phi}^{\mathrm{QT}}(k)\equiv\frac{4\pi k^{3}}{V}\left\langle
\left\vert \phi_{\mathbf{k}}\right\vert ^{2}\right\rangle _{\mathrm{QT}}%
=\frac{H^{2}}{4\pi^{2}}\ .
\end{equation}
The quantity $\left\langle \left\vert \phi_{\mathbf{k}}\right\vert
^{2}\right\rangle _{\mathrm{QT}}$ is calculated from quantum field theory (for
$\lambda_{\mathrm{phys}}>>H^{-1}$). The formula (\ref{form}) then yields a
quantum-theoretical power spectrum%
\begin{equation}
\mathcal{P}_{\mathcal{R}}^{\mathrm{QT}}(k)\equiv\frac{4\pi k^{3}}%
{V}\left\langle \left\vert \mathcal{R}_{\mathbf{k}}\right\vert ^{2}%
\right\rangle _{\mathrm{QT}}=\left[  \frac{H^{2}}{\dot{\phi}_{0}^{2}%
}\mathcal{P}_{\phi}^{\mathrm{QT}}(k)\right]  _{t_{\ast}(k)}=\frac{1}{4\pi^{2}%
}\left[  \frac{H^{4}}{\dot{\phi}_{0}^{2}}\right]  _{t_{\ast}(k)} \label{PRQT}%
\end{equation}
for $\mathcal{R}_{\mathbf{k}}$. In the slow-roll approximation we then obtain
a scale-free spectrum $\mathcal{P}_{\mathcal{R}}^{\mathrm{QT}}(k)\approx
\mathrm{const}.$. Because $H$ and $\dot{\phi}_{0}$ are in fact slowly changing
during the inflationary phase, there will be a small dependence of
$\mathcal{P}_{\mathcal{R}}^{\mathrm{QT}}(k)$ on $k$.

Now quantum nonequilibrium in the early Bunch-Davies vacuum would generally
yield deviations from (\ref{B-D}). It has been shown -- using pilot-wave field
theory on de Sitter space -- that if microscopic quantum nonequilibrium exists
at the onset of inflation then instead of relaxing it will be preserved during
the inflationary phase and then transferred to macroscopic lengthscales by the
spatial expansion \cite{AV07,AV10}.

For each mode, the width of the evolving nonequilibrium distribution maintains
a constant ratio with the width of the equilibrium distribution. This was
shown by calculating the de Broglie-Bohm trajectories for the inflaton field.
Again writing $\phi_{\mathbf{k}}$ in terms of the real quantities
$q_{\mathbf{k}r}$ ($r=1$, $2$), the Bunch-Davies wave functional takes a
product form $\Psi\lbrack q_{\mathbf{k}r},t]=\prod\limits_{\mathbf{k}r}%
\psi_{\mathbf{k}r}(q_{\mathbf{k}r},t)$ where $\left\vert \psi_{\mathbf{k}%
r}\right\vert ^{2}$ is a contracting Gaussian packet of width%
\[
\Delta_{k}(\eta)=\Delta_{k}(0)\sqrt{1+k^{2}\eta^{2}}%
\]
(where $\eta=-1/Ha$ is conformal time, running from $-\infty$ to $0$). In the
late-time limit $\left\vert \psi_{\mathbf{k}r}\right\vert ^{2}$ approaches a
static Gaussian of width $\Delta_{k}(0)=H/\sqrt{2k^{3}}$. Using the de Broglie
equation of motion (\ref{deB2}) it was found that the trajectories take the
form%
\[
q_{\mathbf{k}r}(\eta)=q_{\mathbf{k}r}(0)\sqrt{1+k^{2}\eta^{2}}\ .
\]
From this result one may construct the time evolution of an arbitrary
nonequilibrium distribution $\rho_{\mathbf{k}r}(q_{\mathbf{k}r},\eta)$. It is
readily seen that $\rho_{\mathbf{k}r}$ is a contracting distribution of width%
\[
D_{\mathbf{k}r}(\eta)=D_{\mathbf{k}r}(0)\sqrt{1+k^{2}\eta^{2}}%
\]
(with arbitrary $D_{\mathbf{k}r}(0)$). In the late-time limit $\rho
_{\mathbf{k}r}$ approaches a static packet of width $D_{\mathbf{k}r}(0)$. The
overall time evolution amounts to a homogeneous contraction of both
$\rho_{\mathbf{k}r}$ and $\left\vert \psi_{\mathbf{k}r}\right\vert ^{2}$ by
the same factor. Thus, indeed, for each mode the widths of the nonequilibrium
and equilibrium distributions remain in a fixed ratio over time
\cite{AV07,AV10}.

For simplicity we assume that $D_{\mathbf{k}r}(t)=D_{k}(t)$. We may then write%
\begin{equation}
\frac{D_{k}(t)}{\Delta_{k}(t)}=(\mathrm{const.\ in\ time})\equiv\sqrt{\xi
(k)}\ .
\end{equation}
We then have a nonequilibrium variance%
\begin{equation}
\left\langle |\phi_{\mathbf{k}}|^{2}\right\rangle =\left\langle |\phi
_{\mathbf{k}}|^{2}\right\rangle _{\mathrm{QT}}\xi(k)\ , \label{xi}%
\end{equation}
with a `nonequilibrium function' $\xi(k)\neq1$. The nonequilibrium power
spectrum for $\mathcal{R}_{\mathbf{k}}$ is then%
\begin{equation}
\mathcal{P}_{\mathcal{R}}(k)=\mathcal{P}_{\mathcal{R}}^{\mathrm{QT}}(k)\xi(k)
\end{equation}
and scale invariance is generally broken. Measurements of the angular power
spectrum $C_{l}$ for the CMB may then be used to set experimental bounds on
$\xi(k)$ \cite{AV10}.

To a first approximation we may assume that the quantum-theoretical spectrum
is scale invariant: $\mathcal{P}_{\mathcal{R}}^{\mathrm{QT}}(k)\approx
\mathrm{const}.$. In the low-$l$ region we then have, from (\ref{Clowl}),%
\begin{equation}
\frac{C_{l}}{C_{l}^{\mathrm{QT}}}\approx2l(l+1)\int_{0}^{\infty}\frac{dk}%
{k}\ j_{l}^{2}(2k/H_{0})\xi(k)\ , \label{ratio}%
\end{equation}
where $C_{l}^{\mathrm{QT}}$ denotes the angular power spectrum predicted by
quantum theory and $C_{l}$ denotes that predicted by nonequilibrium pilot-wave
theory. As was pointed out in ref. \cite{AV10}, a low-power anomaly -- that
is, evidence for $C_{l}<C_{l}^{\mathrm{QT}}$ -- may be explained by having
$\xi(k)<1$ in a suitable region of $k$-space.

Note that $\xi(k)<1$ requires that the nonequilibrium width $D_{k}$ for the
inflaton mode be less than the quantum equilibrium width $\Delta_{k}$. It is
reasonable to expect this -- as opposed to $\xi(k)>1$ -- if one accepts our
basic premise that quantum noise has a dynamical origin. For it then seems
natural to assume initial conditions (in this case for a pre-inflationary era)
with a statistical dispersion smaller than the quantum equilibrium value -- so
that the initial state contains less statistical noise than a regular quantum
state. If we make such an assumption, then at later times (as relaxation
proceeds during the pre-inflationary period) the dispersion will reach at most
the equilibrium value. Thus, while a larger-than-quantum inflationary
dispersion ($\xi(k)>1$) is possible in principle, it seems more natural to
have a less-than-quantum dispersion ($\xi(k)<1$).

The integral in (\ref{ratio}) is dominated by the scale $k\approx lH_{0}/2$,
so a significant drop in $C_{l}$ requires $\xi(k)<1$ for $k$ in this region.
Thus we require $\xi(k)<1$ for comoving wavelengths%
\begin{equation}
\lambda\sim(4\pi/l)H_{0}^{-1} \label{cw}%
\end{equation}
that are comparable to the Hubble radius $H_{0}^{-1}$ today.

One might consider a simple cutoff, with $\xi(k)=0$ for $\lambda
>\lambda_{\mathrm{c}}=2\pi/k_{\mathrm{c}}$. The correction to $C_{l}$ will be
significant only if the interval $(0,k_{\mathrm{c}})$ overlaps substantially
with the scale $k\approx lH_{0}/2$, so that $\lambda_{\mathrm{c}}$ cannot be
much larger than $(4\pi/l)H_{0}^{-1}$. If instead we had $\lambda_{\mathrm{c}%
}>>(4\pi/l)H_{0}^{-1}$ the correction to $C_{l}$ would not only be small -- it
would be unobservable (even in principle) because it would be much smaller
than the cosmic variance \cite{AV10}.

To explain the observed power deficit in the low-$l$ region, then, we require
a dip in quantum noise -- quantified by $\xi(k)<1$ -- for modes of wavelength
comparable to (\ref{cw}). Taking $l\lesssim40$, we require a cutoff of order%
\begin{equation}
\lambda_{\mathrm{c}}\sim H_{0}^{-1}\ . \label{cutoff}%
\end{equation}

\subsection{Infra-red cutoff $\lambda_{\mathrm{c}}$ from a pre-inflationary
era}

There are likely to be many possible mechanisms for producing such a cutoff.
One scenario might involve a suitable period of `fast rolling' for the
inflaton field \cite{CL03}. Another scenario, outlined here, would involve a
radiation-dominated pre-inflationary era with suppression of quantum noise at
large scales.

Let us consider a radiation-dominated pre-inflationary phase starting at an
initial time $t_{i}$, with a transition to an inflationary phase occurring
around a time $t_{f}$. As shown in Figure 5, the transition from pre-inflation
to inflation is modelled (for simplicity) as a sudden jump at time $t_{f}$. It
will be necessary to assume that $H^{-1}$ increases across the jump, from
$H_{-}^{-1}(t_{f})$ to $H_{+}^{-1}(t_{f})>H_{-}^{-1}(t_{f})$.

Let us denote the approximately constant Hubble radius during inflation by
$H_{\mathrm{\inf}}^{-1}$ (equal to $H_{+}^{-1}(t_{f})$). Relevant cosmological
fluctuations -- those that make a measurable contribution to the CMB --
originate from inside $H_{\mathrm{\inf}}^{-1}$. If some of those modes were
out of equilibrium during inflation, they must have evolved from modes that
did not completely relax during the pre-inflationary phase (where the
pre-inflationary modes are understood to refer to the relevant fields that
were then present). Given our results for relaxation on expanding space with
$a\propto t^{1/2}$, relic nonequilibrium at the end of pre-inflation is most
likely to exist for modes that remained in the super-Hubble regime.

We therefore focus our attention on field modes that enter the Hubble radius
during the transition from pre-inflation to inflation. As shown in Figure 5,
for such modes no time is spent in the (pre-inflationary) sub-Hubble regime
and therefore relaxation is likely to be suppressed. However, relaxation could
still occur during the transition itself, around the time $t_{f}$. If we
assume that nonequilibrium can survive the transition, then these modes can
still be out of equilibrium at the beginning of inflation and make a
nonequilibrium contribution to the CMB spectrum -- provided $H_{-}^{-1}%
(t_{f})<H_{\inf}^{-1}$, so that modes outside the Hubble radius just before
$t_{f}$ can be inside the Hubble radius just after $t_{f}$.%

%TCIMACRO{\FRAME{ftbpFU}{4.8291in}{3.4832in}{0pt}{\Qcb{Inflation with a
%radiation-dominated pre-inflationary era. The dashed line shows the Hubble
%radius $H^{-1}$. The solid lines show physical wavelength $\lambda
%_{\QTR{rm}{phys}}$ for two different modes: the lower line enters the Hubble
%radius during pre-inflation and exits during inflation, while the upper line
%remains outside the Hubble radius throughout the pre-inflationary era and
%enters only during the transition.}}{}{figv.jpg}%
%width 4.8291in;  height 3.4832in;  depth 0pt;  original-width 10.6665in;
%original-height 7.6795in;  cropleft "0";  croptop "1";  cropright "1";
%cropbottom "0";  filename 'FigV.jpg';file-properties "XNPEU";}} }%
%BeginExpansion
\begin{figure}
\begin{center}
\includegraphics[width=0.8\textwidth]%
{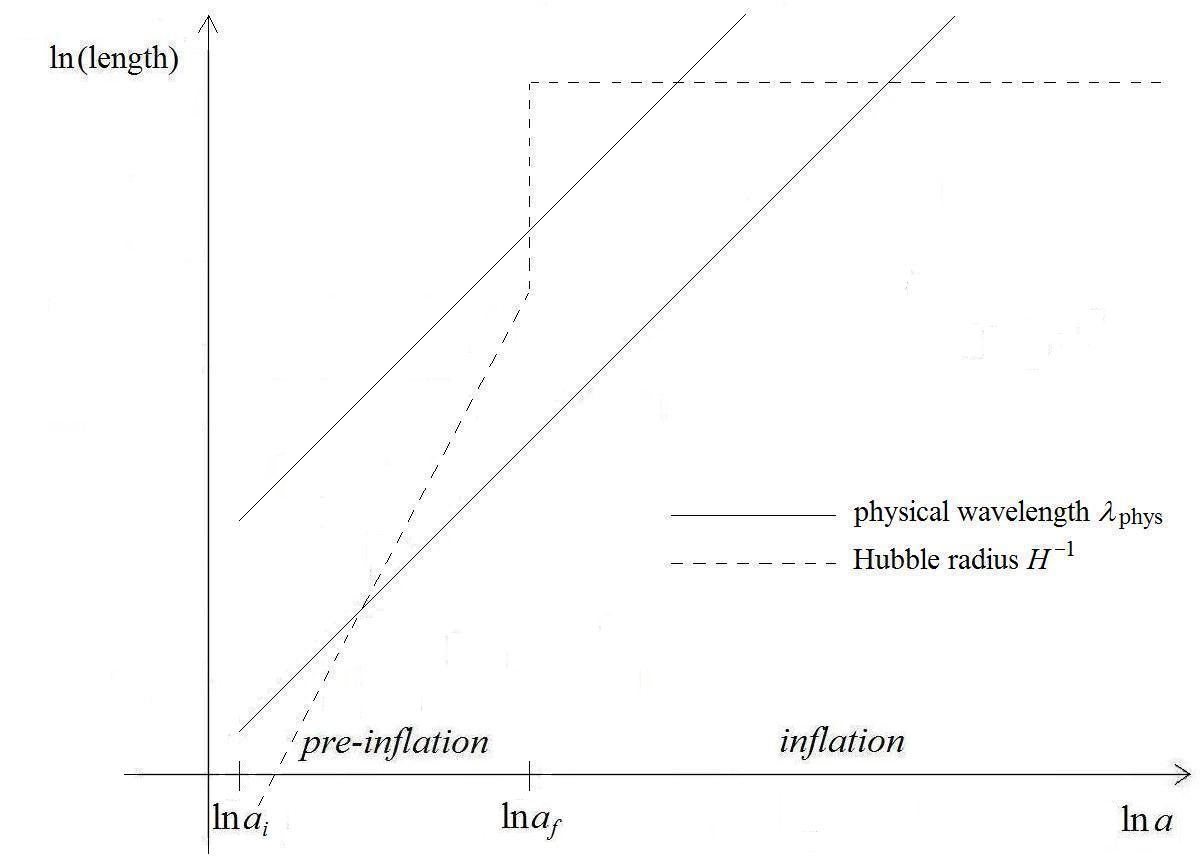}%
\caption{Inflation with a radiation-dominated pre-inflationary era. The dashed
line shows the Hubble radius $H^{-1}$. The solid lines show physical
wavelength $\lambda_{\mathrm{phys}}$ for two different modes: the lower line
enters the Hubble radius during pre-inflation and exits during inflation,
while the upper line remains outside the Hubble radius throughout the
pre-inflationary era and enters only during the transition.}%
\end{center}
\end{figure}
%EndExpansion

Modes can enter the Hubble radius only if $\lambda_{\mathrm{phys}}$ increases
more slowly than does $H^{-1}$ -- that is, only if the comoving Hubble radius
$h^{-1}\equiv H^{-1}/a=1/\dot{a}$ increases with time. This occurs for a
decelerating universe ($\ddot{a}<0$), which requires that the pressure $p$ and
energy density $\rho$ satisfy $w\equiv p/\rho>-1/3$. If our putative
nonequilibrium modes are to contribute to the CMB spectrum, $h^{-1}$ must
increase during the transition from pre-inflation to inflation. To show that
this could occur, let us consider how $h^{-1}$ varies as a function of $a$.
Writing%
\[
dh^{-1}/da=(1/\dot{a})dh^{-1}/dt=-(h^{-1})^{3}\ddot{a}%
\]
and using the Friedmann--Lema\^{\i}tre equations%
\[
\frac{\ddot{a}}{a}=-\frac{4\pi G}{3}(\rho+3p)\ ,
\]%
\[
\left(  \frac{\dot{a}}{a}\right)  ^{2}=\frac{8\pi G}{3}\rho
\]
yields%
\[
\frac{d\ln h^{-1}}{da}=\frac{u}{a}\ ,
\]
where the parameter%
\[
u\equiv\frac{1}{2}(1+3w)
\]
varies from $+1$ to $-1$ as the equation-of-state parameter $w$ varies from
$+1/3$ to $-1$. We may then integrate across the transition, yielding%
\begin{equation}
\frac{h_{2}^{-1}}{h_{1}^{-1}}=\exp\left(  \int_{a_{1}}^{a_{2}}\frac{u}%
{a}da\right)  \label{result}%
\end{equation}
(where subscripts $1$ and $2$ denote values at the beginning and end of the
transition respectively). We will have the desired increase, $h_{2}^{-1}%
/h_{1}^{-1}>1$, if and only if%
\begin{equation}
\int_{a_{1}}^{a_{2}}\frac{u}{a}da>0\ .
\end{equation}
Because $u/a$ ranges from $1/a_{1}$ to $-1/a_{2}$, where $a_{2}>a_{1}$, it is
plausible that this integral will indeed be positive (though logarithmically
small) -- in which case physical wavelengths will be driven inside the Hubble
radius, thereby allowing the said modes to contribute to the CMB spectrum.

A proper discussion of the transition would require a detailed model, and it
is quite possible that relaxation -- or at least significant relaxation --
will occur during the transition. On the other hand, the transition takes
place from a pre-inflationary era of relaxation suppression for super-Hubble
modes to an inflationary era of totally suppressed relaxation on all scales.
It then seems possible that nonequilibrium modes that are outside the Hubble
radius just before the transition will not completely relax during the
transition. Here we shall simply assume that if nonequilibrium exists
immediately prior to $t_{f}$ then it will survive, at least to some degree,
until the beginning of inflation itself. (A future strategy to model the
transition is noted in Section 8.)

If we make that assumption, then nonequilibrium is possible for all modes such
that $\lambda_{\mathrm{phys}}(t_{f})\gtrsim H_{-}^{-1}(t_{f})$. We may then
obtain an estimate for the cutoff $\lambda_{\mathrm{c}}$ -- the minimal
comoving wavelength for which nonequilibrium is likely to exist -- by setting%
\begin{equation}
a_{f}\lambda_{\mathrm{c}}\sim H_{-}^{-1}(t_{f})\ . \label{lb2}%
\end{equation}

The scale factor $a_{f}$ (at the end of pre-inflation) may be written as%
\[
a_{f}=a_{f}/a_{0}=(a_{f}/a_{\mathrm{end}})(a_{\mathrm{end}}/a_{0})\ ,
\]
where $a_{\mathrm{end}}$ is the scale factor at the end of inflation. The
expansion that takes place during the transition from pre-inflation to
inflation may presumably be neglected compared to the huge expansion that
takes place during inflation itself. We may then approximately identify
$a_{f}$ with the scale factor $a_{\mathrm{begin}}$ at the beginning of
inflation -- in which case we have $a_{f}/a_{\mathrm{end}}\simeq e^{-N}$,
where $N$ is the number of inflationary e-folds. If we similarly neglect the
expansion that takes place during the transition from inflation to
post-inflation, we can write $a_{\mathrm{end}}/a_{0}\simeq T_{0}%
/T_{\mathrm{end}}$ (where $T_{\mathrm{end}}$ is the temperature at which
inflation ends). Thus we have%
\begin{equation}
a_{f}\simeq e^{-N}(T_{0}/T_{\mathrm{end}})
\end{equation}
and so we find%
\begin{equation}
\lambda_{\mathrm{c}}\sim e^{N}H_{-}^{-1}(t_{f})\left(  T_{\mathrm{end}}%
/T_{0}\right)  \ .
\end{equation}

Since (inserting $c$, and using the standard temperature clock $t\sim
(1\ \mathrm{s})\left(  1\ \mathrm{MeV}/k_{\mathrm{B}}T\right)  ^{2}$ for a
radiation-dominated era)%
\[
H_{-}^{-1}(t_{f})=2ct_{f}\sim2c(1\ \mathrm{s})\left(  1\ \mathrm{MeV}%
/k_{\mathrm{B}}T_{f}\right)  ^{2}\sim(10^{11}\ \mathrm{cm})\left(
1\ \mathrm{MeV}/k_{\mathrm{B}}T_{f}\right)  ^{2}\ ,
\]
and using $k_{\mathrm{B}}T_{0}\sim10^{-4}\ \mathrm{eV}$, we find%
\[
\lambda_{\mathrm{c}}\sim(10^{-1}\ \mathrm{cm})e^{N}\left(  T_{\mathrm{end}%
}/T_{f}\right)  \left(  T_{\mathrm{P}}/T_{f}\right)  \ ,
\]
where $T_{\mathrm{P}}$ is the Planck temperature. Writing $(1\ \mathrm{cm}%
)\simeq H_{0}^{-1}e^{-65}$ (where $H_{0}^{-1}\simeq10^{28}\ \mathrm{cm}$), we
have an approximate formula%
\begin{equation}
\lambda_{\mathrm{c}}\sim10^{-1}H_{0}^{-1}e^{(N-65)}\left(  T_{\mathrm{end}%
}/T_{f}\right)  \left(  T_{\mathrm{P}}/T_{f}\right)  \label{cutoff4}%
\end{equation}
for the cutoff $\lambda_{\mathrm{c}}$ in terms of three parameters $N$,
$T_{\mathrm{end}}/T_{f}$ and $T_{\mathrm{P}}/T_{f}$.

This is of course only a rough estimate. Even so, because (\ref{cutoff4}) was
derived from essentially `kinematical' arguments we may expect that the true
expression for the cutoff will not be strongly model-dependent and that
(\ref{cutoff4}) will provide an indication of the order of magnitude. On the
other hand, of course, the actual values of the parameters appearing in
(\ref{cutoff4}) will be strongly model-dependent.\footnote{One could also
consider modes that enter the Hubble radius during pre-inflation (cf. Figure
5), but do not spend enough time in the sub-Hubble regime for them to relax
completely. Consideration of these modes yields a small correction to the
expression (\ref{cutoff4}) for the cutoff \cite{AVbook}.}

One may reasonably expect $T_{f}$ to be of the same order of magnitude as the
energy scale $H_{\mathrm{\inf}}\sim10^{16}\ \mathrm{GeV}\sim10^{-3}%
T_{\mathrm{P}}$ associated with the inflationary phase. Thus we may take%
\begin{equation}
T_{\mathrm{P}}/T_{f}\sim10^{3}\ .
\end{equation}
Our estimate (\ref{cutoff4}) for $\lambda_{\mathrm{c}}$ is then%
\begin{equation}
\lambda_{\mathrm{c}}\sim10^{2}H_{0}^{-1}e^{(N-65)}\left(  T_{\mathrm{end}%
}/T_{f}\right)  \ . \label{cutoff5}%
\end{equation}
We have two parameters: the number $N$ of e-folds and the `reheating ratio'
$T_{\mathrm{end}}/T_{f}$.

For inflation to solve the horizon and flatness problems, standard estimates
indicate that the minimum number $N=N_{\min}$ of e-folds required -- from the
beginning of inflation to the end of inflation -- is $N_{\min}\simeq70$
(though some authors take $N_{\min}\simeq60$). See, for example, ref.
\cite{PU09}. (It is of course possible that the actual number $N$ of e-folds
is much larger than $N_{\min}$. See, for example, ref. \cite{M04}.)

The ratio $T_{\mathrm{end}}/T_{f}$ depends on the details of the reheating
process. If the inflaton decay time is smaller than the Hubble time (evaluated
at the end of inflation), the vacuum energy is expected to be rapidly
converted into radiation. The predicted `reheating temperature'
$T_{\mathrm{end}}$ depends, among other things, on the inflaton decay rate.
Estimates for $T_{\mathrm{end}}/T_{f}$ depend on the model, and can range from
$T_{\mathrm{end}}/T_{f}\sim1$ to $T_{\mathrm{end}}/T_{f}<<1$. (For overviews
of the theory of reheating see, for example, refs. \cite{PU09,Alla10}.) One
may also attempt to constrain $T_{\mathrm{end}}$ by means of CMB data
\cite{MR10,M11}. Martin and Ringeval \cite{MR10} obtain lower bounds on
$T_{\mathrm{end}}$ in the range $390\ \mathrm{GeV}-890\ \mathrm{TeV}$
(depending on the inflationary model), corresponding to lower bounds on
$T_{\mathrm{end}}/T_{f}$ in the range $\sim10^{-14}-10^{-10}$ (assuming
$T_{f}\sim10^{-3}T_{\mathrm{P}}$).

For the estimate (\ref{cutoff5}) to yield a $\lambda_{\mathrm{c}}$ of the
required order of magnitude (\ref{cutoff}), we have the constraint%
\begin{equation}
e^{(N-65)}\left(  T_{\mathrm{end}}/T_{f}\right)  \sim10^{-2}\ .
\end{equation}
This is consistent with the allowed parameter space. For example, we could
have $N\sim65$ and $T_{\mathrm{end}}/T_{f}\sim10^{-2}$. To have much more than
the minimal number of e-folds requires a very small reheating ratio. For
example, if we allow $T_{\mathrm{end}}/T_{f}$ to be as small as $\sim10^{-10}$
then $N$ can range up to $\sim83$.

If instead $e^{(N-65)}\left(  T_{\mathrm{end}}/T_{f}\right)  >>10^{-2}$ then
$\lambda_{\mathrm{c}}>>H_{0}^{-1}$ and the angular power spectrum for low $l$
will be unaffected. In this case the nonequilibrium (even if it exists) will
be completely unobservable. There are of course models of inflation in which
$N>>65$. For these, there would be no hope of detecting pre-inflationary
nonequilibrium in the CMB. However, such models do lead to an alternative
possibility: when $N>>N_{\min}$ it can happen that the Hubble radius today
originated from a lengthscale which, at the beginning of inflation, was
smaller than the Planck length \cite{BM01,MB01}. Such `trans-Planckian' modes
could be subject to novel gravitational effects that generate quantum
nonequilibrium \cite{AV04,AV07}, yielding an observable effect on the CMB
\cite{AV10}. A small value of $N$, therefore, makes it more likely that we
could detect pre-inflationary nonequilibrium in the CMB; while a large value
of $N$ makes it more plausible that we could detect a Planck-scale production
of nonequilibrium in the CMB (if such effects exist). There might be
intermediate values of $N$ such that neither effect would be visible. Only
further and more detailed model building can tell us where such intermediate
values may lie.

\section{Conclusion}

We have constructed an exactly-solvable model for the suppression of quantum
noise at super-Hubble wavelengths in a radiation-dominated universe. The
results broadly confirm expectations of a suppression of relaxation to quantum
equilibrium for super-Hubble modes \cite{AV07,AV08a,AV10,AVbook}. The
mechanism emerges naturally from pilot-wave dynamics on expanding space. We
have also considered a cosmological scenario with a pre-inflationary phase, to
illustrate how the mechanism might explain the large-scale power deficit that
has recently been confirmed to exist in the CMB \cite{PlanckXV}.

The statistical significance of the observed low-$l$ power deficit is not
high: it could be a random fluctuation for our single sky, as opposed to a
genuine anomaly in the underlying power spectrum. A better understanding of
the deficit and of its significance requires the development of physical
models that (i) predict such a deficit, and (ii) make additional testable
predictions. The first requirement has been met by showing that the de
Broglie-Bohm pilot-wave theory contains a natural mechanism for producing a
suppression of quantum noise at large scales on expanding space. The second
requirement is a matter for future work -- some suggestions will be made here.

Firstly, it would be of interest to study the detailed application of our
mechanism for quantum noise suppression to specific cosmological models, with
a view to predicting features of the nonequilibrium function $\xi(k)$ that
modifies the inflationary power spectrum. In our scenario with a
pre-inflationary phase, for example, one could study the evolution of
nonequilibrium in the pre-inflationary era -- including across the transition
to inflation. This would require a model of the transition. Given the scale
factor $a=a(t)$ as a function of time during the transition, it should be
possible to solve the ordinary differential equations (\ref{eqdiff1}%
)--(\ref{eqdiff3}) (at least numerically) and thereby obtain the wave
functional for the field modes. One could then calculate the de Broglie-Bohm
trajectories and study how early nonequilibrium evolves across the transition.
Given a prediction for $\xi(k)$, one could then make a comparison with current
data -- and weigh the outcome against rival explanations based on other models.

Secondly, in this paper we have focussed for definiteness on a scenario with
quantum noise suppression in a radiation-dominated expansion. It is however
conceivable that a pre-inflationary phase was not radiation-dominated. Our
solution for the wave functional may be readily generalised to an expansion
with a power law $a\propto t^{p}$ since, as is well known, the mode equation
(\ref{classical_shoes}) may then be solved in terms of Bessel functions. (The
phase factor $\Theta$ would still be given by the integral (\ref{Theta}).) One
could then investigate quantum noise suppression for more general spatial expansions.

Thirdly, we note that quantum nonequilibrium in the inflationary phase can
generate non-Gaussianity, which can manifest as non-random phases and
inter-mode correlations \cite{AV10}. An early relaxation suppression could
certainly generate non-Gaussian effects, though this remains to be studied in
detail. While some authors have suggested that non-Gaussianity may exist in
the \textit{WMAP} data \cite{YW08}, little evidence for it has so far been
found in the \textit{Planck} data \cite{PlanckXXIII,PlanckXXIV}.

Finally, the \textit{Planck} team has also reported tentative evidence for
anisotropy at large scales \cite{PlanckXXIII}. Whether or not this feature is
truly primordial remains to be seen. It is in any case worth noting that a
large-scale anisotropy could be generated by quantum nonequilibrium simply by
allowing the width $D_{\mathbf{k}r}(t)$ of the (nonequilibrium) inflaton field
distribution $\rho_{\mathbf{k}r}(q_{\mathbf{k}r},t)$ to depend on the
direction of the mode wave vector $\mathbf{k}$ -- and not just on its
magnitude $k$ as was assumed in ref. \cite{AV10}. We would then have a
nonequilibrium function $\xi=\xi(\mathbf{k})$ that depends on the direction of
$\mathbf{k}$. If such a non-isotropic nonequilibrium existed in a
pre-inflationary phase, isotropy would be recovered in the inflationary era
for those modes that relaxed to equilibrium -- since the inflationary
equilibrium width $\Delta_{k}(t)$ depends only on $\left\vert \mathbf{k}%
\right\vert $ (and $t$). On the other hand, if relaxation suppression occurs
for long-wavelength modes during pre-inflation then the anisotropy will
presumably remain at large scales during inflation -- along with the power
deficit. We therefore seem to have a single mechanism whereby both a power
deficit and a statistical anisotropy can be generated at large angular scales
in the CMB. Whether such a scenario could provide a good fit to the
\textit{Planck} data is left for future analysis.

\textbf{Acknowledgements}. AV wishes to thank Patrick Peter for helpful
discussions and comments on the manuscript. This research was funded jointly
by the John Templeton Foundation and Clemson University.

\end{document}